\documentclass[reprint,onecolumn,
amsmath,amssymb,aps]{revtex4}
\usepackage{epsfig,bm,epsf,graphics}
\usepackage{graphicx}
\usepackage{dcolumn}
\usepackage{bm}
\usepackage{xcolor}
\begin{document}
\draft
\preprint{APS/123-QED}
\title{QUANTUM SPIN-WAVE THEORY FOR NON-COLLINEAR SPIN STRUCTURES, A REVIEW}
\author{H. T. Diep\footnote{diep@cyu.fr}}
\affiliation{%
Laboratoire de Physique Th\'eorique et Mod\'elisation,
CY Cergy Paris Universit\'e, CNRS, UMR 8089\\
2, Avenue Adolphe Chauvin, 95302 Cergy-Pontoise Cedex, France.\\
 }%

\date{\today}

\begin{abstract}
In this review, we trace the evolution of the quantum spin-wave theory treating  non-collinear spin configurations.  Non-collinear spin configurations are consequences of the frustration created by competing interactions. They include simple chiral magnets due to competing nearest-neighbor (NN) and next-NN interactions and systems with geometry frustration such as the triangular antiferromagnet and the Kagom\'e lattice. We review here spin-wave results of such systems and also systems with the Dzyaloshinskii-Moriya interaction.  Accent is put on these non-collinear ground states which have to be calculated before applying any spin-wave theory to determine the spectrum of the elementary excitations from the ground states. We mostly show results obtained by the use of a  Green's function method. These results include the spin-wave dispersion relation and the magnetizations, layer by layer, as functions of  $T$ in 2D, 3D and  thin films. Some new unpublished results are also included. Technical details and discussion on the method are shown and discussed.
\vspace{0.5cm}
\begin{description}
\item PACS numbers: 75.25.-j ; 75.30.Ds ; 75.70.-i \\
\item Keywords: Quantum Spin-Wave Theory; Green's Function Theory; Frustrated Spin Systems ; Non-Collinear Spin Configurations; Dzyaloshinskii-Moriya Interaction; Phase Transition; Monte Carlo Simulation.
\end{description}
\end{abstract}


\maketitle

\section{Introduction}

In a solid the interaction between its constituent atoms or molecules gives rise to elementary excitations from its ground state (GS) when the temperature increases from zero. One has examples of elementary excitations due to atom-atom interaction known as phonons or due to spin-spin interaction known as magnons. Note that magnons are spin waves (SW) when they are quantized.  Elementary excitations are defined also for interaction between charge densities in plasma, or for electric dipole-dipole interaction in ferroelectrics, among others. Elementary excitations are thus collective motions which dominate the low-temperature behaviors of solids in general.

For a given system, there are several ways to calculate the energy of elementary excitations from classical treatments to quantum ones. Since those collective motions are waves, its energy depends on the wave vector $\mathbf k$. The $\mathbf k$-dependent energy is often called the SW spectrum for spin systems.  Note that though the calculation of the SW spectrum is often for periodic crystalline structures, it can also be performed for symmetry-reduced systems such as in thin films or in semi-infinite solids in which the translation symmetry is broken by the presence of a surface.

In this review we focus on the SW excitations in magnetically ordered systems. The history began with ferromagnets and antiferromagnets with collinear spin GSs, parallel or antiparallel configurations in the early 50's. Most of the works on the SW used either the classical method or the quantum Holstein-Primakoff transformation.  The Green's function (GF) technique has also been introduced in a pioneer paper of Zubarev \cite{Zubarev}. The first application of this method to thin films has been done  \cite{DiepGF1979}. Note that unlike the SW theory, the GF can treat the SW up to higher temperatures. We will come back to this point later.

Let us recall some important breakthroughs in the study of  non-collinear spin configurations. The first discovery of the helical spin configuration has been published in 1959 \cite{Yoshimori,Villain1959}. Some attempts to treat this non-collinear case have been done in the 70's and 80's. Let us cite two noticeable works on this subject in Refs.  \cite{Rastelli1985,DiepSW1989}. In these works, a local system of spin coordinates have been introduced in the way that each spin lies on its quantization axis. One can therefore use the commutation relations between spin deviation operators. These works took into account magnon-magnon interactions by expanding the Hamiltonian up to three-operator terms at temperature $T=0$ \cite{Rastelli1985} or up to four-operator terms at low $T$ \cite{DiepSW1989}. Nevertheless, since these works used the Holstein-Primakoff method, the case of higher $T$ cannot be dealt with.  In Ref. \cite{Quartu1997}, the GF method has been employed for  the first time to calculate the SW spectrum in a frustrated system where the GS spin configuration is non collinear. Using the SW spectrum, the local order parameter, the specific heat, ... were calculated.  Since this work, we have applied the GF method to a variety of systems where the  GS is non collinear. In this review, we will recall results of some of these published works.

Let us comment on the frustration which is the origin of the non-collinear GS.  The frustration is caused by  either the competing interactions in the system or a geometry frustration as in the  triangular lattice with only the antiferromagnetic interaction between the  nearest neighbors (NN) (see Ref. \cite{DiepFSS}). The frustration causes high GS degeneracy, and for the vector spins ( XY and Heisenberg cases) the  spin configurations are non collinear making the calculation of the SW spectrum harder. A number of examples will be shown in this review paper.

In addition to competing interactions, the Dzyaloshinskii-Moriya (DM) interaction \cite{Dzyaloshinskii,Moriya} is also the origin of non-collinear spin configurations in spin systems. While the Heisenberg model between two spins is written as $-J_{ij}\mathbf S_i\cdot \mathbf S_j$ giving rise to two collinear spins in the GS, the DM interaction is written as $\mathbf D_{ij}\cdot \mathbf S_i \times \mathbf S_j$ giving rise to two perpendicular spins. The DM model was historically proposed to explain the phenomenon of weak ferromagnetism observed in Mn compounds \cite{Sergienko}. However, the DM interaction is at present  discoververed in various materials, in particular at the interface of a multilayer \cite{Stashkevich,Heide,Ederer,Cepas,Rohart}. Although in this review we do not show the effect of the DM interaction in a magnetic field which gives rise to topological spin swirls known as skyrmions, we should mention a few of the important works given in Refs. \cite{Bogdanov,Muhlbauer,Yu2,Seki}. Skyrmions are among the most studied subjects at the time being due to their potental applications in spinelectronics.\cite{Fert2013} We refer the reader to the  rich biography given in our recent papers in Refs. \cite{Zhang2020,Zhang2021}.

Since this paper is a review on the method and the results of published works on SW in non-collinear GS spin configurations, it is important to recall the method and show main results of some typical cases.  We would like to emphasize that on the GF technique, to our knowledge there are no authors other than us working with this method. Therefore, the works mentioned in the references of this paper are our works published over the last 25 years.
The aim of this review is two-fold. First we show technical details of the GF method by selecting a number of subjects which are of current interest in research: helimagnets, systems including a DM interaction, surface effects in thin films. Second, we show that these systems possess many striking features due to the frustration.

This paper is organized as follows. In section \ref{heli}, we express the Hamiltonian in a general non-collinear GS  and define the local system of spin coordinates. Here, we also present the calculation of the GS and the foundation of the self-consistent GF technique and the calculation of the SW dispersion relation and layer magnetizations at arbitrary temperature ($T$). We show in section \ref{heliresult} the numerical results obtained from the GF.  Section \ref{DMI} shows interesting examples using various kinds of interaction including the DM interaction in a variety of systems from two dimensions, to thin films and superlattices. Section \ref{DMAFTL} treats a case where the DM interaction competes with the antiferromagnetic interaction in the frustrated antiferromagnetic triangular lattice.  Section \ref{surface} presents the surface effect in a thin film where its surface is frustrated.
Concluding remarks are given in section \ref{concl}.

\section{Hamiltonian of a Chiral Magnet - Local Coordinates}\label{heli}

Chiral order in helimagnets has been subject of recent extensive investigations.  In Ref. \cite{Mello2003}, the surface structure of thin helimagnetic films has been studied. In Ref. \cite {Cinti2008} exotic spin configurations in ultrathin helimagnetic holmium films have been investigated. In Refs. \cite{Karhu2011,Karhu2012}
chiral structure and spin reorientations in MnSi thin films have been theoretically studied.  In these works, the chiral structures have been considered at $T=0$, but not the SW even at $T=0$. The main difficulty was due to the non-collinear, non-uniform spin configurations. We have shown that this was possible using the GFs generalized for such spin configurations given in Ref. \cite{Quartu1997}

\begin{figure}[ht!]
\centering
\includegraphics[width=6cm,angle=0]{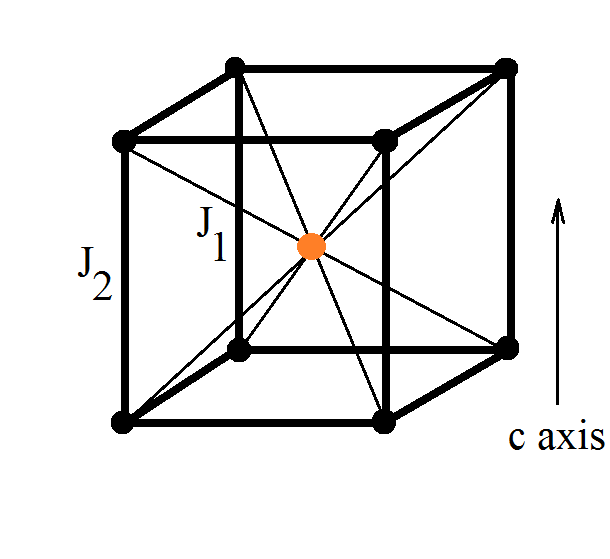}
\caption{Interactions $J_1$ (thin solid lines) between nearest neighbors and $J_2$ between next nearest neighbors along the $c$ axis in a bct lattice.\label{bctfig}}
\end{figure}

To demonstrate the method, let us follow Ref. \cite{Diep2015}: we consider the body-centered tetragonal (bct) lattice with Heisenberg spins. Each spin interacts with its nearest neighbors (NN) via the exchange constant $J_1$ and with its next NN (NNN) on the $c$-direction via the exchange $J_2$ (see Fig.\ref{bctfig}).

We consider the simplest model of helimagnet given by the following Hamiltonian
\begin{equation}\label{eqn:hamil1}
{\cal H}=-J_1\sum_{i,j} \mathbf S_i \cdot \mathbf S_j -J_2\sum_{i,k} \mathbf S_i \cdot \mathbf S_k
\end{equation}
where $\mathbf S_i$ is a quantum spin of magnitude 1/2,  the first sum is performed over all NN pairs and the second sum over pairs on the $c$-axis (cf. Fig. \ref{bctfig}).

In the case of an infinite crystal, the chiral state occurs when $J_1$ is ferromagnetic and $J_2$ is
antiferromagnetic and $|J_2|/J_1$ is larger than a critical value, as will be shown below.

Let us suppose that the energy of a spin $E_C$ in a chiral configuration when the angle between two NN spin in the neighboring planes is $\theta$, one has (omitting the factor $S^2$)
\begin{equation}\label{energy}
E= -8J_1\cos \theta-2J_2\cos (2\theta)
\end{equation}
The lowest-energy state corresponds to
\begin{eqnarray}
\frac {dE}{d\theta}&=&0 \nonumber\\
\rightarrow  8J_1\sin\theta+4\sin(2\theta)&=&0\nonumber\\
8J_1\sin\theta(1+\frac{J_2}{J_1}cos\theta)&=&0
\end{eqnarray}
There are two solutions,
\noindent
$\sin \theta=0$ and $\cos\theta=-\frac{J_1}{J_2}$
\noindent
The first solution corresponds to the ferromagnetic state, and the second solution exists if  $-\frac{J_1}{J_2}\leq 1$ which corresponds to the chiral state.

\begin{figure}[htb]
\centering
\includegraphics[width=5cm]{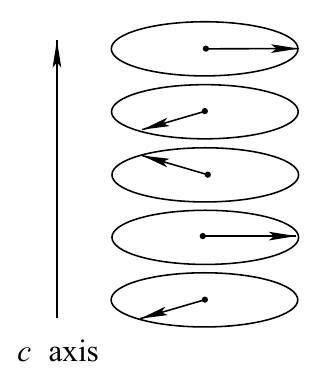}  
\includegraphics[width=7cm,angle=0]{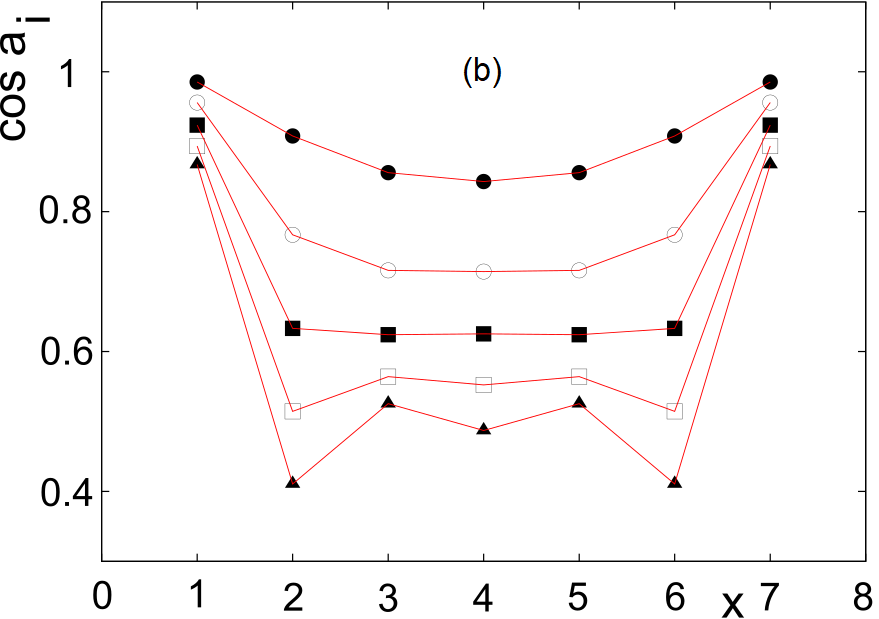}  
\caption{ (a) Chiral structure along the $c$-axis for an infinite crystal,
in the case $\theta=2\pi/3$, namely $J_2/J_1=-2$; (b) Cosinus of $\alpha_1=\theta_1-\theta_2$, ..., $\alpha_7=\theta_7-\theta_8$ across the film
for several values $J_2/J_1=-1.2,-1.4,-1.6,-1.8, -2$ (from top) with $N_z=8$: $a_i$ stands for $\theta_i-\theta_{i+1}$ and $x$ indicates the film layer $i$ where the angle $a_i$ with the layer $(i+1)$ is shown.  See text for comments. }\label{angle}
\end{figure}

For a thin helimagnetic film, the angle between spins in adjacent layers varies due to the surface. We can use the method of energy minimization for each layer, then we have a set of coupled equations to solve (see Ref. \cite{Diep2015}). Figure \ref{angle} displays an example of the angle distribution across the film thickness $N_z$.

\begin{figure}[ht!]
\centering
\includegraphics[width=6cm,angle=0]{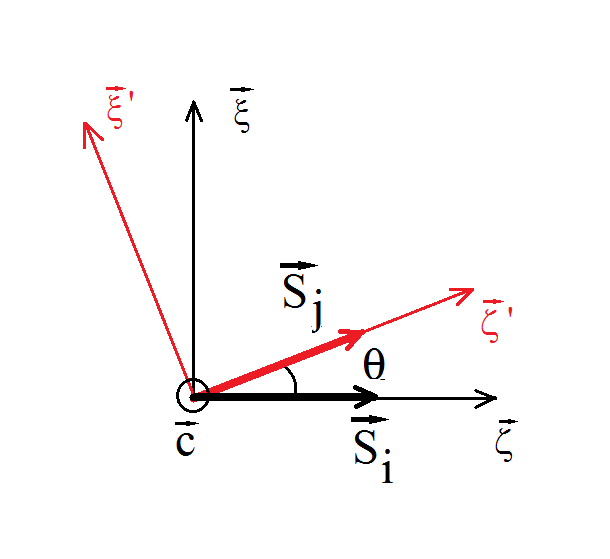}
\caption{Spin $\mathbf S_i$ lies along the $\vec \zeta$ axis (its quantization axis), while spin  $\mathbf S_j$ lies along its quantization axis $\vec \zeta'$ which makes an angle $\theta$ with the $\vec \zeta$ axis.  The axes $\vec \xi$ and $\vec \xi'$ are perpendicular respectively to $\vec \zeta$ and $\vec \zeta'$. T he perpendicular axes $\hat \eta_i$ and $\hat \eta_j$ coincide with the $\vec c$ axis, perpendicular to the basal plane of the bct lattice.}\label{localco}
\end{figure}

In order to calculate the SW spectrum for systems of non-collinear spin configurations, let us emphasize that the commutation relations between spin operators are established when the spin lies on its quantization $z$.  In the non-collinear cases, each spin has its own quantization axis. It is therefore important to choose a quantization axis for each spin. We have to use the system of local coordinates defined as follows. In the Hamiltonian, the spins are coupled two by two. Consider a pair $\mathbf S_i$ and $\mathbf S_j$. As seen above,  in the general case these spins make an angle $\theta_{i,j}=\theta_j-\theta_i$ determined by the competing interactions in the systems. For quantum spins, in the course of calculation we need to use the commutation relations between the spin operators $S^z,S^+,S^-$. As said above, these commutation relations are derived from the assumption that the spin lies on  its quantization axis $z$.
We show in Fig. \ref{localco} the local coordinates assigned to spin $\mathbf S_i$ and $\mathbf S_j$. We write

\begin{eqnarray}
\mathbf S_i&=&S_i^x\hat \xi_i+S_i^y\hat \eta_i+S_i^z\hat \zeta_i\label{SI}\\
\mathbf S_j&=&S_j^x\hat \xi_j+S_j^y\hat \eta_j+S_j^z\hat \zeta_j\label{SJ}
\end{eqnarray}
Expressing the axes of $\mathbf S_j$ in the frame of $\mathbf S_i$ one has
\begin{eqnarray}
\hat \zeta_j&=&\cos \theta_{i,j}\hat \zeta_i+\sin \theta_{i,j}\hat \xi_i\label{localco2}\\
\hat \xi_j&=&-\sin \theta_{i,j}\hat \zeta_i+\cos \theta_{i,j}\hat \xi_i\label{localco1}\\
\hat \eta_j&=&\hat \eta_i\label{localco3}
\end{eqnarray}
so that
\begin{eqnarray}
\mathbf S_{j}&=&S_j^x(-\sin\theta_{i,j}\hat \zeta_i +\cos \theta_{i,j}\hat \xi_i)\nonumber\\
&&+S_j^y\hat \eta_i+S_j^z(\cos\theta_{i,j}\hat \zeta_i +\sin \theta_{i,j}\hat \xi_i)\label{local6}
\end{eqnarray}

Using Eq. (\ref{local6}) to express $\mathbf S_j$ in the $(\hat \xi_i,\hat \eta_i,\hat \zeta_i)$ coordinates, we  calculate
$\mathbf S_i \cdot \mathbf S_j$, we get the following Hamiltonian from (\ref{eqn:hamil1}):
\begin{eqnarray}
\mathcal H_e &=& - \sum_{<i,j>}
J_{i,j}\Bigg\{\frac{1}{4}\left(\cos\theta_{i,j} -1\right)
\left(S^+_iS^+_j +S^-_iS^-_j\right)\nonumber\\
&+& \frac{1}{4}\left(\cos\theta_{i,j} +1\right) \left(S^+_iS^-_j
+S^-_iS^+_j\right)\nonumber\\
&+&\frac{1}{2}\sin\theta_{i,j}\left(S^+_i +S^-_i\right)S^z_j
-\frac{1}{2}\sin\theta_{i,j}S^z_i\left(S^+_j
+S^-_j\right)\nonumber\\
&+&\cos\theta_{i,j}S^z_iS^z_j\Bigg\}
\label{eq:HGH2}
\end{eqnarray}
This explicit Hamiltonian in terms of the angle between two NN spins is common for a non-collinear spin configuration
due to exchange interactions $J_{i,j}$. For other types of interactions such as the DM interaction, the explicit Hamiltonian in terms of the angle will be different as shown
in section \ref{DMI}.

We define the following GFs for the above Hamiltonian:

\begin{eqnarray}
G_{i,j}(t,t')&=&<<S_i^+(t);S_{j}^-(t')>>\nonumber\\
&=&-i\theta (t-t')
<\left[S_i^+(t),S_{j}^-(t')\right]> \label{green59a}\\
F_{i,j}(t,t')&=&<<S_i^-(t);S_{j}^-(t')>>\nonumber\\
&=&-i\theta (t-t')
<\left[S_i^-(t),S_{j}^-(t')\right]>\label{green60}
\end{eqnarray}
Writing their equations of motion we have
\begin{eqnarray}
i\hbar\frac {d}{dt}G_{i,j}\left( t,t'\right) &=& \left<\left[ S^+_i
\left( t\right) , S^-_j \left( t'\right)\right]\right>\delta\left(
t-t'\right) \nonumber\\
&-& \left<\left< \left[\mathcal H, S^+_i\left( t\right)\right] ;
S^-_j \left( t'\right) \right>\right>,
\label{eq:HGEoMG}\\
i\hbar\frac {d}{dt}F_{i,j}\left( t,t'\right) &=& \left<\left[ S^-_i
\left( t\right) , S^-_j \left( t'\right)\right]\right>\delta\left(
t-t'\right)\nonumber \\
&-& \left<\left< \left[\mathcal H, S^-_i\left( t\right)\right] ;
S^-_j \left( t'\right) \right>\right>, \label{eq:HGEoMF}
\end{eqnarray}
where
\begin{eqnarray}
S^{\pm}_j&=&S^x_j\hat \xi_j \pm i S^y_j\hat \eta_j\nonumber\\
\left [S^+_j,S^-_l \right ]&=&2S^z_j\delta_{j,l} \nonumber\\
\left [S^{z}_j,S^{\pm}_l \right ]&=&\pm S^{\pm}_j\delta_{j,l}\nonumber
\end{eqnarray}
Note that the equation of motion of the G Green's function generates the F Green's functions, and vice-versa.
Performing the commutators in Eqs. (\ref{eq:HGEoMG})-(\ref{eq:HGEoMF}), and using the Tyablikov approximation \cite{Tyablikov} for
higher-order GFs, for instance $<<S_{i'}^zS_i^+(t);S_{j}^-(t')>>\simeq <S_{i'}^z><<S_i^+(t);S_{j}^-(t')>>$ etc., we obtain

 \begin{eqnarray}
 i\hbar \frac{dG_{i,j}(t,t')}{dt}&=&2<S_i^z>\delta_{i,j} \delta (t-t')\nonumber\\
 &-&\sum_{i'}J_{i,i'}[<S_i^z>(\cos \theta_{i,i'}-1)\times \nonumber\\
 &\times& F_{i',j}(t,t')\nonumber\\
&+&<S_i^z>(\cos \theta_{i,i'}+1)G_{i',j}(t,t')\nonumber\\
&-&2<S_{i'}^z>\cos \theta_{i,i'}G_{i,j}(t,t')]\nonumber\\
&+&2\sum_{i'}I_{i,i'}<S_{i'}^z>\cos \theta_{i,i'}G_{i,j}(t,t')\nonumber\\
&&\label{GFG0}\\
i\hbar \frac{dF_{i,j}(t,t')}{dt}&=&\sum_{i'}J_{i,i'}[<S_i^z>(\cos \theta_{i,i'}-1)\times\nonumber\\
&\times& G_{i',j}(t,t')\nonumber\\
&+&<S_i^z>(\cos \theta_{i,i'}+1)F_{i',j}(t,t')\nonumber\\
&-&2<S_{i'}^z>\cos \theta_{i,i'}F_{i,j}(t,t')]\nonumber\\
&-&2\sum_{i'}I_{i,i'}<S_{i'}^z>\cos \theta_{i,i'}F_{i,j}(t,t')\nonumber\\
&&\label{GFF0}
\end{eqnarray}
Note that the Tyablikov decoupling scheme is equivalent to the so-called "random-phase-approximation" (RPA).

For the sake of clarity, we write separately the NN and NNN sums, we have
 \begin{eqnarray}
 i\hbar \frac{dG_{i,j}(t,t')}{dt}&=&2<S_i^z>\delta_{i,j} \delta (t-t')\nonumber\\
 &-&\sum_{k'\in NN}J_{i,k'}[<S_i^z>(\cos \theta_{i,k'}-1) \times \nonumber\\
 &\times& F_{k',j}(t,t')\nonumber\\
 &+&<S_i^z>(\cos \theta_{i,k'}+1) G_{k',j}(t,t')\nonumber\\
 &-&2<S_{k'}^z>\cos \theta_{i,k'}G_{i,j}(t,t')]\nonumber\\
 &+&2\sum_{k'\in NN}I_{i,k'}<S_{k'}^z>\cos \theta_{i,k'}G_{i,j}(t,t')\nonumber\\
&-&\sum_{i'\in NNN}J_{i,i'}[<S_i^z>(\cos \theta_{i,i'}-1) \times \nonumber\\
&\times& F_{i',j}(t,t')\nonumber\\
&+&<S_i^z>(\cos \theta_{i,i'}+1) G_{i',j}(t,t')\nonumber\\
&-&2<S_{i'}^z>\cos \theta_{i,i'}G_{i,j}(t,t')]\label{GFG2}\\
i\hbar \frac{dF_{k,j}(t,t')}{dt}&=&\sum_{i'\in NN}J_{k,i'}[<S_k^z>(\cos \theta_{k,i'}-1)\times\nonumber\\
&\times&  G_{i',j}(t,t')\nonumber\\
&+&<S_k^z>(\cos \theta_{k,i'}+1) F_{i',j}(t,t')\nonumber\\
&-&2<S_{i'}^z>\cos \theta_{k,i'}F_{k,j}(t,t')]\nonumber\\
&-&2\sum_{i'\in NN}I_{k,i'}<S_{i'}^z>\cos \theta_{k,i'}F_{k,j}(t,t')\nonumber\\
&+&\sum_{k'\in NNN}J_{k,k'}[<S_k^z>(\cos \theta_{k,k'}-1) \times \nonumber\\
&\times& G_{k',j}(t,t')\nonumber\\
&+&<S_k^z>(\cos \theta_{k,k'}+1) F_{k',j}(t,t')\nonumber\\
&-&2<S_{k'}^z>\cos \theta_{k,k'}F_{k,j}(t,t')]\label{GFF2}
\end{eqnarray}
For simplicity, we suppose in the following  $J_{k,k'}$ are all equal to $J_1$ for NN interactions and to $J_2$ for NNN interactions.
 $I_{k,k'}$ is taken to be $I_1$ for NN pairs.  In addition, in the film coordinates defined above, we denote the Cartesian components of the spin position $\mathbf R_i$ by three indices $(\ell_i,m_i,n_i)$ in three directions $x$, $y$ and $z$.

Since there is the translation invariance in the $xy$ plane, the in-plane Fourier
transforms of the above equations in the $xy$ plane are

\begin{eqnarray}
G_{i, j}\left( t, t'\right) &=& \frac {1}{\Delta}\int\int_{BZ} d\mathbf
k_{xy}\frac{1}{2\pi}\int^{+\infty}_{-\infty}d\omega e^{-i\omega
\left(t-t'\right)}\nonumber\\
&&\hspace{0.7cm}\times g_{n_i,n_j}\left(\omega , \mathbf k_{xy}\right)
e^{i\mathbf k_{xy}\cdot \left(\mathbf R_i-\mathbf
R_j\right)},\label{eq:HGFourG}\\
F_{k, j}\left( t, t'\right) &=& \frac {1}{\Delta}\int\int_{BZ} d\mathbf
k_{xy}\frac{1}{2\pi}\int^{+\infty}_{-\infty}d\omega e^{-i\omega
\left(t-t'\right)}\nonumber\\
&&\hspace{0.7cm}\times f_{n_k,n_j}\left(\omega , \mathbf k_{xy}\right)
e^{i\mathbf k_{xy}\cdot \left(\mathbf R_k-\mathbf
R_j\right)},\label{eq:HGFourF}
\end{eqnarray}
where $\omega$ is the SW frequency, $\mathbf k_{xy}$
the wave-vector parallel to $xy$ planes and $\mathbf R_i$
the position of  $\mathbf S_i$. $n_i$, $n_j$ and $n_k$ denote
the $z$-components of the sites $\mathbf R_i$,  $\mathbf R_j$ and $\mathbf R_k$. The integral over $\mathbf k_{xy}$ is performed in the
first Brillouin zone ($BZ$) whose surface is $\Delta$ in the $xy$
reciprocal plane.  $n_i=1$ denotes the surface layer, $n_i=2$ the second layer etc.

In the 3D case, the Fourier transformation  of Eqs. (\ref{GFG2})-(\ref{GFF2}) in the three $(x,y,z)$ directions yields the SW spectrum in the absence of anisotropy:
\begin{equation}
\hbar\omega=\pm \sqrt{A^2-B^2}
\end{equation}
where
\begin{eqnarray}
A&=& J_1 \left< S^z\right>[\cos\theta+1]Z\gamma+2Z J_1\left< S^z\right> \cos \theta\nonumber\\
&&+J_2 \left< S^z\right> [\cos (2\theta)+1]Z_c\cos (k_za) \nonumber\\
&&+2Z_cJ_2 \left< S^z\right> \cos (2\theta)\nonumber\\
B&=& J_1 \left< S^z\right> (\cos\theta -1)Z\gamma\nonumber\\
&&+J_2 \left< S^z\right>[\cos(2\theta) -1]Z_c\cos (k_za) \nonumber
\end{eqnarray}
where $Z=8$ is the NN coordination number, $Z_c=2$  the NNN number on the $c$-axis and $\gamma=\cos (k_xa/2)\cos (k_ya/2)\cos (k_za/2)$  where $a$ is the lattice constant taken the same in three directions.
Note that $\hbar\omega$ is zero when $A=\pm B$. This is realized at two points as expected in helimagnets: $k_x=k_y=k_z=0$ ($\gamma=1$) and  $k_z=2\theta$ along the helical axis.
It is interesting to note that we recover the SW dispersion relation of ferromagnets (antiferromagnets)  \cite{DiepGF1979} with NN interaction only by putting $\cos \theta=1$ $(-1)$ in the above coefficients.

In the case of a thin film, the in-plane Fourier transformation yields the following matrix equation
\begin{equation}
\mathbf M \left( \omega \right) \mathbf h = \mathbf u,
\label{eq:HGMatrix}
\end{equation}
where $\mathbf h$ and $\mathbf u$ are
given by
\begin{equation}
\mathbf h = \left(%
\begin{array}{c}
  g_{1,n'} \\
  f_{1,n'} \\
  \vdots \\
  g_{n,n'} \\
  f_{n,n'} \\
    \vdots \\
  g_{N_z,n'} \\
  f_{N_z,n'} \\
\end{array}%
\right) , \mathbf u =\left(%
\begin{array}{c}
  2 \left< S^z_1\right>\delta_{1,n'}\\
  0 \\
  \vdots \\
  2 \left< S^z_{N_z}\right>\delta_{N_z,n'}\\
  0 \\
\end{array}%
\right) , \label{eq:HGMatrixguH}
\end{equation}
We take $\hbar=1$ hereafter. Note that $\mathbf M\left(\omega\right)$ is a
$\left(2N_z \times 2N_z\right)$ matrix given by Eq. (\ref{eq:HGMatrixMH})
\begin{equation}\label{eq:HGMatrixMH}
\mathbf M\left(\omega\right) = \left(%
\begin{array}{cccccccccccc}
  \omega+A_1&0    & B^+_1    & C^+_1& D_1^+& E_1^+& 0&0&0&0&0&0\\
   0    & \omega-A_1  & -C^+_1 & -B^+_1 &-E_1^+&-D_1^+&0&0&0&0&0&0\\
   \cdots & \cdots & \cdots &\cdots&\cdots&\cdots&\cdots&\cdots&\cdots&\cdots&\cdots&\cdots\\
 \cdots&D_n^-&E_n^-&B^-_{n}&C^-_{n}&\omega+A_{n}&0&B^+_{n}&C^+_{n}&D_n^+&E_n^+&\cdots\\
 \cdots&-E_n^-&-D_n^-&-C^-_{n}&-B^-_{n}&0&\omega-A_{n}&-C^+_{n}&-B^+_{n}&-E_n^+&-D_n^+&\cdots\\
         \cdots  & \cdots & \cdots & \cdots &\cdots&\cdots&\cdots&\cdots&\cdots&\cdots&\cdots&\cdots \\
  0& 0&0&0& 0& 0& D^-_{N_z}& E^-_{N_z}  & B^-_{N_z}   & C^-_{N_z}   &\omega + A_{N_z}&0\\
  0&0&0&0&0&0&-E^-_{N_z}& -D^-_{N_z} & -C^-_{N_z}  & -B^-_{N_z}&0  & \omega-A_{N_z}\\
\end{array}%
\right)
\end{equation}
where
\begin{eqnarray}
A_{n} &=& - 8J_1(1+d) \Big[\left< S^z_{n+1}\right>
\cos\theta_{n,n+1}\nonumber\\
&&+\left< S^z_{n-1}\right>
\cos\theta_{n,n-1}\Big]\nonumber\\
&-&2J_2 \Big[\left< S^z_{n+2}\right>
\cos\theta_{n,n+2}\nonumber\\
&&+\left< S^z_{n-2}\right>
\cos\theta_{n,n-2}\Big]\nonumber
\end{eqnarray}

\begin{eqnarray}
B_n^\pm &=& 4J_1 \left< S^z_{n}\right>(\cos\theta_{n,n\pm 1}+1)\gamma \nonumber\\
C_n^\pm &=& 4J_1 \left< S^z_{n}\right>(\cos\theta_{n,n\pm 1}-1)\gamma \nonumber\\
E_n^\pm &=& J_2 \left< S^z_{n}\right>(\cos\theta_{n,n\pm 2}-1)\nonumber\\
D_n^\pm &=& J_2 \left< S^z_{n}\right>(\cos\theta_{n,n\pm 2}+1) \nonumber
\end{eqnarray}
where we recall that $n$ denotes the layer number, namely $1,2,...,N_z$ and $d=I_1/J_1$. Note that $\theta_{n,n\pm
1}$ denotes the angle between a spin in the layer $n$ and its NN spins in adjacent layers $n\pm 1$ etc. and
$\gamma = \cos \left( \frac{k_x a}{2} \right)\cos \left( \frac{k_y a}{2} \right).$

In order to obtain the SW fequency $\omega$, we solve the secular equation $\det|\mathbf M|=0$ for each given ($k_x,k_y)$.  Since the linear dimension of the square marix is 2$N_z$, we obtain  2$N_z$ eigen-values of $\omega$, half positive and half negative,  corresponding to two opposite spin precessions as in antiferromagnets.  These values depend on the input values $<S_n^z>$ ($n=1,...,N_z$).  Thus, we have to solve the secular equation by iteration until the convergence of input and output values.
Note that  even at $T=0$,  $<S_n^z>$ are not equal to $1/2$ due to the zero-point spin contraction \cite{DiepTM}. In addition, because of the film surfaces, the spin contractions are not uniform.

The solution for $g_{n,n}$ can be calculated (see Ref.  \cite{Diep2015}).
The spectral theorem \cite{Zubarev} can be used to
obtain, after a somewhat lengthy algebra (see \cite{Diep2015}),:
\begin{equation}
\langle S_{n}^z\rangle=\frac{1}{2}-
   \frac{1}{\Delta}
   \int
   \int dk_xdk_y
   \sum_{i=1}^{2N_z}\frac{D_{2n-1}(\omega_i)}
   {\mbox{e}^{\beta \omega_i}-1}
\end{equation}\label{lm2H}
where $n=1,...,N_z$, and
\begin{equation}
D_{2n-1}\left(\omega_i\left(\mathbf k_{xy}\right)\right) = \frac{\left|
\mathbf M\right|_{2n-1} \left(\omega_i\left(\mathbf
k_{xy}\right)\right)}{\prod_{j\neq i}\left[\omega_j\left(\mathbf
k_{xy}\right)-\omega_i\left(\mathbf k_{xy}\right)\right]}.
\end{equation}
As $<S_{n}^z>$ depends each other in $\omega_i (i=1,...,2N_z)$,
their solutions  should be obtained by iteration
at a given temperature $T$. In the particular case where $T=0$ one has

\begin{equation}\label{surf38}
\langle S_{n}^z\rangle(T=0)=\frac{1}{2}+
   \frac{1}{\Delta} \int \int dk_xdk_y
   \sum_{i=1}^{N_z}D_{2n-1}(\omega_i\left(\mathbf k_{xy}\right))
\end{equation}
Note that the sum is performed over  $N_z$ negative $\omega_i$ since for positive $\omega_i$ yield the zero Bose-Einstein factor at $T=0$).

The transition temperature $T_c$ can be calculated  self-consistently when all  $<S_{n}^z>$  tend to zero.

We show in the following section, the numerical results using the above formulas.

\section{Results for helimagnets obtained from the Green's function technique}\label{heliresult}
We use the ferromagnetic interaction between NN as unit, namely $J_1=1$.  Take the helimagnetic case where  $J_2$ is negative with
$|J_2|>J_1$. We have determined above the spin configuration across the film for several values of $p=J_2/J_1$.
Replacing the angles $\theta_{n,n\pm1}$ and $\theta_{n,n\pm2}$ in the matrix elements of $\left|\mathbf M\right|$, then calculating  $\omega_i(i=1,...,2N_z)$ for each $\mathbf k_{xy}$. For the iterative procedure, the reader is referred to Re. \cite{Diep2015}. The solution
 $\langle S_{n}^z\rangle (n=1,...,N_z)$ is obtained when the input and the output are equal with a desired precision $P$.

\subsection{Spectrum}

\begin{figure}[htb]
\centering
\includegraphics[width=7cm,angle=0]{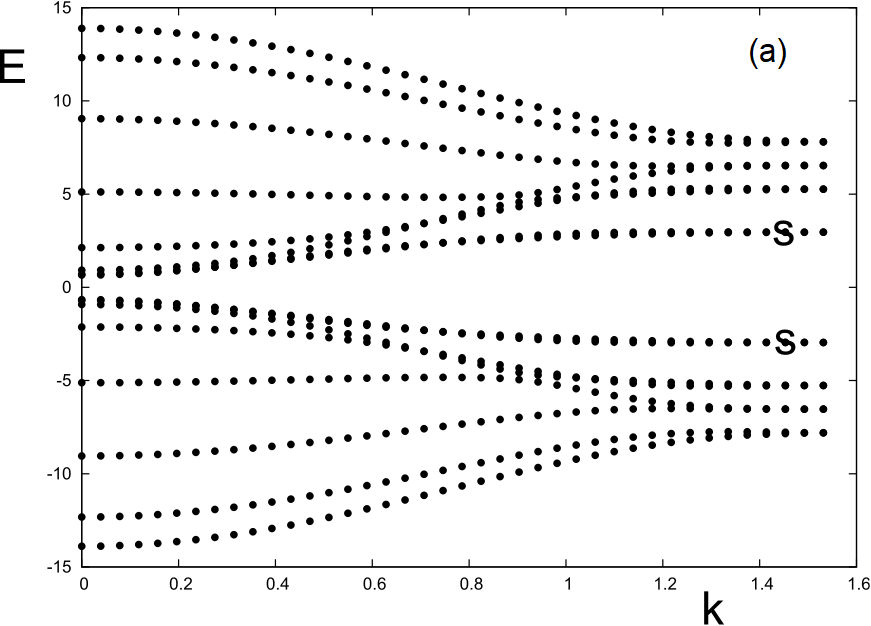}  
\includegraphics[width=7cm,angle=0]{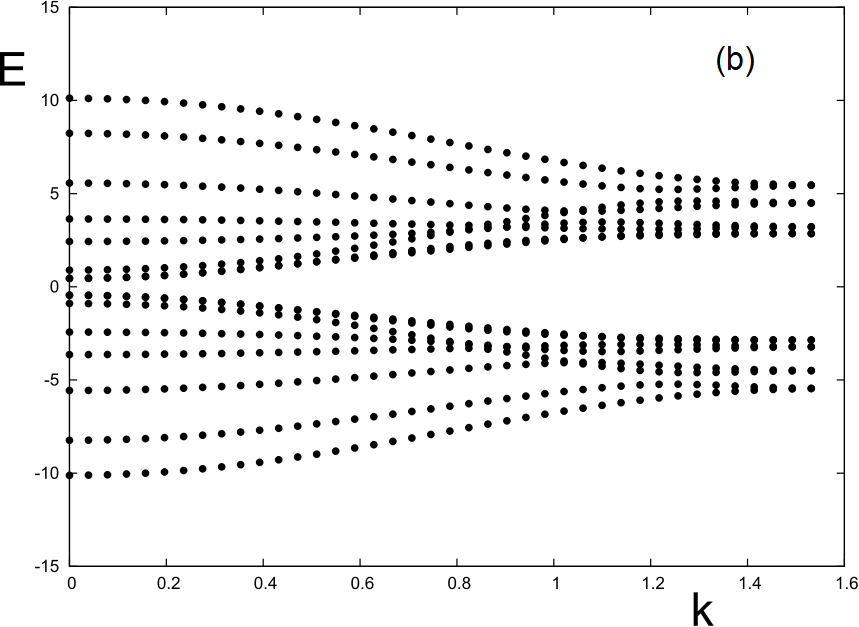}  
\caption{ (a) Spectrum $E=\hbar \omega$ versus $k\equiv k_x=k_y$ for $J_2/J_1=-1.4$ at $T=0.1$  and (b) $T=1.02$, for $N_z=8$ and $d=0.1$. The surface branches are indicated by $s$.}\label{sweta}
\end{figure}

We calculate the SW spectrum as described above for each a given $J_2/J_1$. The SW spectrum depends on $T$. We show in Fig. \ref{sweta} the SW spectrum  $\omega$ versus $k_x=k_y$ for an 8-layer film with $J_2/J_1=-1.4$  at $T=0.1$ and $T=1.02$ (in units of $J_1/k_B=1$).   We observe that

\noindent (i) There are opposite-precession SW modes. Unlike ferromagnets, SW in antiferromagnets and non collinear spin structures have opposite spin precessions \cite{DiepTM}. The negative sign does not mean SW negative energy, but it indicates just the precession contrary to the trigonometric sense, \\
(ii) There are two degenerate acoustic "surface" branches one on each side. These degenerate "surface" modes stem from the symmetry of the two surfaces. These surface modes propagate parallel to the film surface but are damped when going to the bulk,\\
(iii) With increasing $T$, layer magnetizations decrease as seen hereafter, this reduces therefore the SW frequency (see Fig. \ref{sweta}b),\\
(iv)  Surface and bulk SW spectra have been observed by inelastic neutron scattering in collinear magnets (ferro- and antiferromagnetic films) \cite{Heinrich,Zangwill}.  However,  such experiments have not been reported for helimagnetic thin films.

\subsection{Zero-point spin contraction and transition temperature}
\begin{figure}[htb]
\centering
\includegraphics[width=7cm,angle=0]{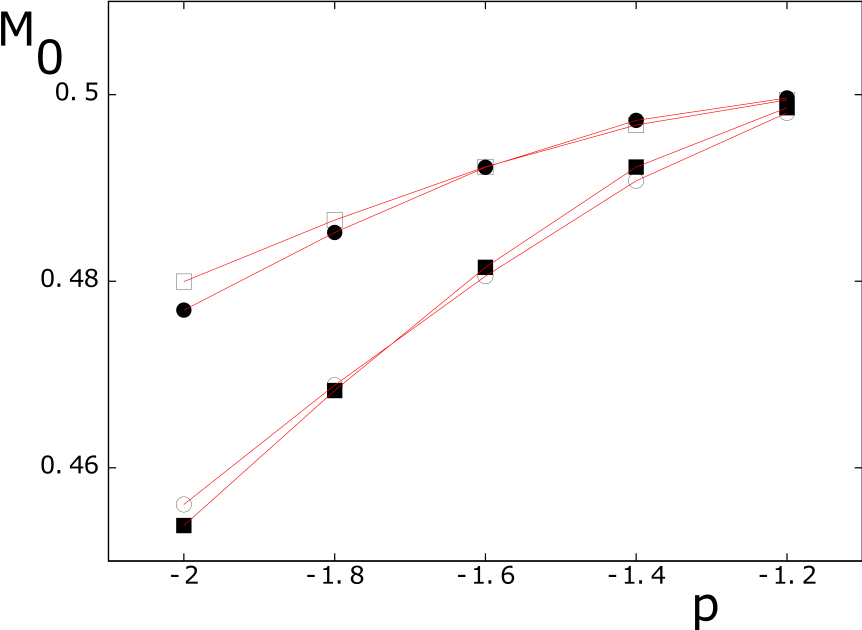}  
\caption{Spin lengths of the first four layers at $T=0$ for several values of $p=J_2/J_1$ with $d=0.1$, $N_z=8$. As seen, all spins are contracted to values smaller than the spin magnitude 1/2.
Black circles,  void circles, black squares and void squares are for first, second, third and fourth layers, respectively.  }\label{spin0}
\end{figure}

Ii isknown that in antiferromagnetic materials, quantum fluctuations cause a contraction of the spin length, namely the spin length is shorter than the spin magnitude, at $T=0$ \cite{DiepTM}.  We demonstrate here that a spin with a stronger antiferromagnetic interaction has a stronger contraction: spins in the first and in the second layers have only one antiferromagnetic NNN on the $c$-axis while interior spins have two NNN.  The contraction at a given $J_2/J_1$ is thus expected to be stronger for interior spins. This is shown in Fig. \ref{spin0}: with increasing $|J_2|/J_1$, i.e. the antiferromagnetic interaction becomes stronger,  the contraction is stronger. Of course,  there is no contraction when the system is ferromagnetic, namely when $J_2 \rightarrow -1$.

\subsection{Layer magnetizations}

\begin{figure}[htb]
\centering
\includegraphics[width=7cm,angle=0]{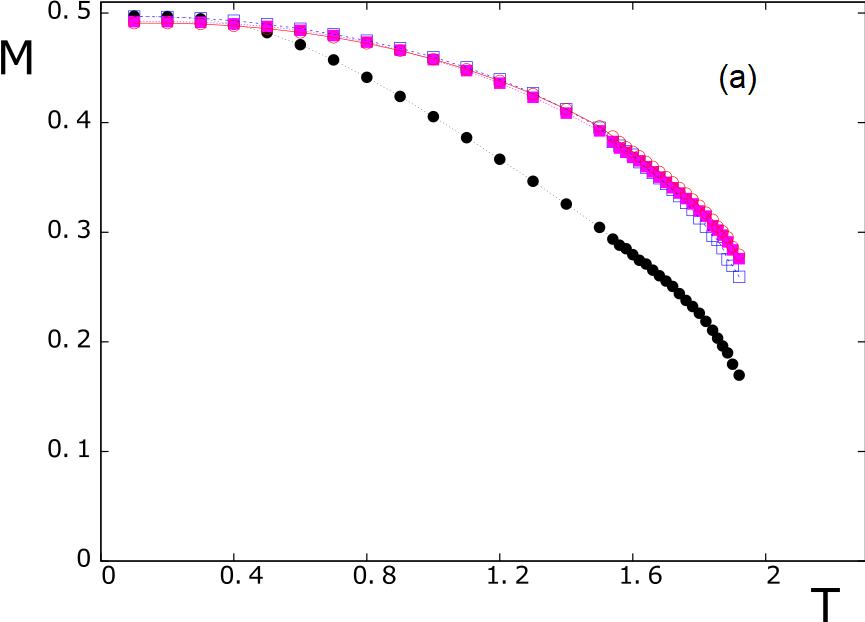}  
\includegraphics[width=7cm,angle=0]{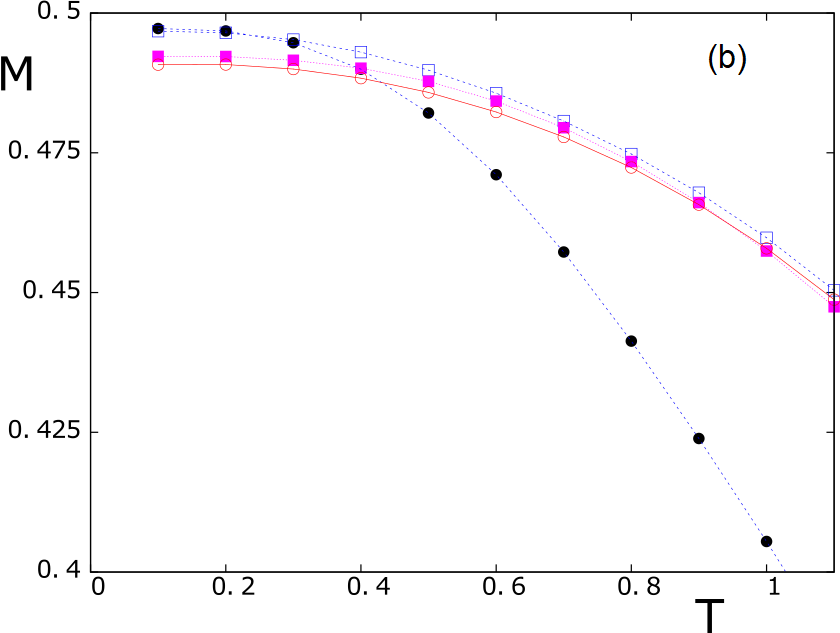}  
\caption{ (a) Layer magnetizations as functions of $T$ for $J_2/J_1=-1.4$ with $d=0.1$, $N_z=8$, (b) Zoom of the region at low $T$ to show crossover. Black circles, blue void squares, magenta squares and red void circles are for first, second, third and fourth layers, respectively.  See text.}\label{magnet14}
\end{figure}

\begin{figure}[htb]
\centering
\includegraphics[width=7cm,angle=0]{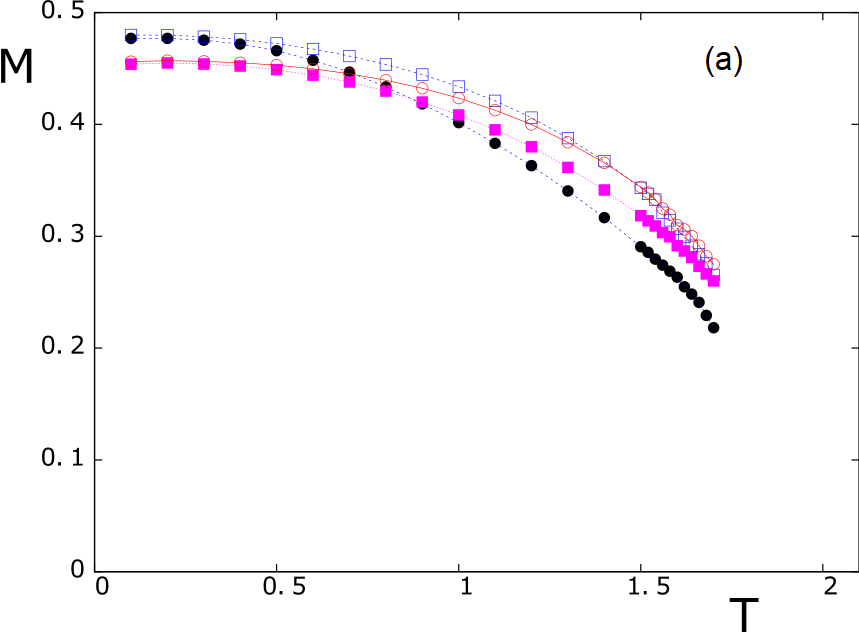}  
\includegraphics[width=7cm,angle=0]{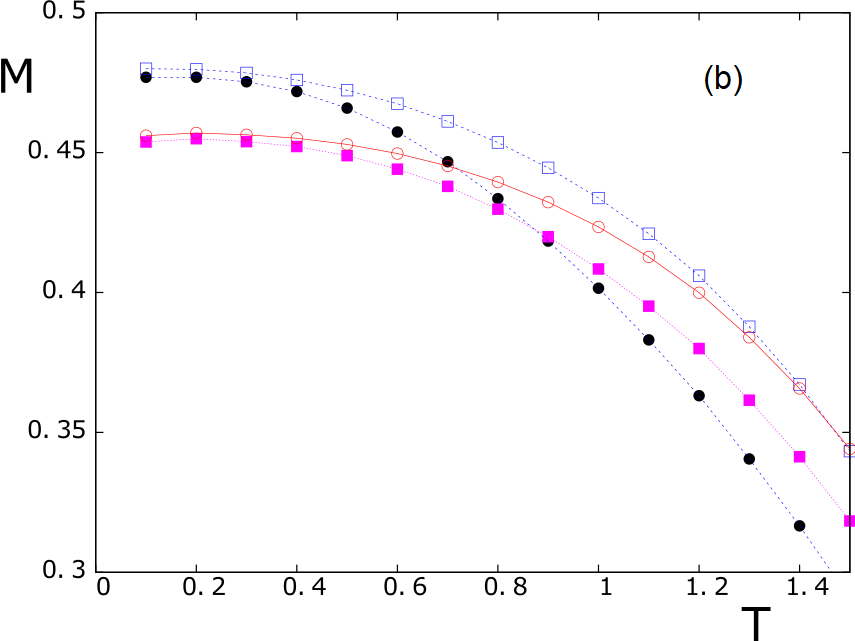}  
\caption{ (a) Layer magnetizations as functions of $T$ for $J_2/J_1=-2$ with $d=0.1$, $N_z=8$, (b) Zoom of the region at low $T$ to show crossover. Black circles, blue void squares, magenta squares and red void circles are for first, second, third and fourth layers, respectively. See text.}\label{magnet20}
\end{figure}

We show now the layer ordering in Figs. \ref{magnet14} and \ref{magnet20} where  $J_2/J_1=-1.4$ and -2, respectively, in the case of $N_z=8$.   Consider first the case $J_2/J_1=-1.4$. We note that
the surface magnetization, having a large value at $T=0$ as seen in Fig. \ref{spin0}, crosses the interior layer magnetizations at $T\simeq 0.42$ to become much smaller than interior magnetizations at higher temperatures.  This crossover phenomenon is due to the competition between quantum fluctuations, which dominate low-$T$ behavior, and the low-lying surface SW modes which reduce the surface magnetization at higher $T$.  Note that the second-layer magnetization makes also a crossover at $T\simeq 1.3$ which is more complicated to analyze.  Similar crossovers have been observed in other quantum systems such as antiferromagnetic films \cite{DiepTF91} and superlattices \cite{DiepSL89}.
Similar remarks are also hold for $J_2/J_1=-2$ shown in Fig. \ref{magnet20}.


Note that the results shown above have been calculated with an in-plane anisotropy interaction $d=0.1$.  Larger $d$ yields stronger layer magnetizations and larger $T_c$.

To close this section on SW in helimagnetic bct thin films, we mention that a same investigation was done in the case of simple-cubic helimagnetic films where the surface spin reconstruction and the surface SW have been shown. \cite{Sahbi} We have also studied the frustrated bct Heisenberg helimagnet in which the SW spectrum of the non-collinear spin configuration has been calculated.\cite{QuartuJMMM1998bctheli}

\section{Dzyaloshinskii-Moriya interaction in thin films}\label{DMI}

Let us consider a thin film made of $N$ square lattices stacked in
the $y$ direction perpendicular to the film surface.  The results for this system have been published in Ref. \cite{Diep2017}.
Hereafter, we review some of these  important results.
The Hamiltonian is given by

\begin{eqnarray}
\mathcal H&=&\mathcal H_e+\mathcal H_{DM}\label{eqn:hamil1}\\
\mathcal H_e&=&-\sum_{\left<i,j\right>}J_{i,j}\mathbf S_i\cdot\mathbf S_j\label{eqn:hamil2}\\
\mathcal H_{DM}&=&\sum_{\left<i,j\right>}\mathbf D_{i,j}\cdot \mathbf S_i\times\mathbf S_j
 \label{eqn:hamil3}
\end{eqnarray}
where $J_{i,j}$ and $\mathbf D_{i,j}$ are the exchange and DM interactions, respectively,
between two quantum Heisenberg spins $\mathbf S_i$ and $\mathbf S_j$ of magnitude $S=1/2$.

We supppose in this section the in-plane and inter-plane exchange interactions between NN
are both ferromagnetic and denoted by $J_1$ and $J_2$, respectively.
The DM interaction is defined only between NN in the plane for simplicity.
The $J$ term favors the collinear spin configuration while the
DM term favors the perpendicular one, this will lead to a compromise where  $\mathbf S_i$ makes
an angle $\theta_{i,j}$ with its neighbor $\mathbf S_j$.
It is obvious that  the quantization axes of $\mathbf S_i$ and $\mathbf S_j$ are different.
Therefore, the transformation using the local coordinates, Eqs. (\ref{SI})-(\ref{local6}), is necessary.
Let us suppose that the vector $\mathbf D_{i,j}$ is along the $y$ axis, namely the $\hat\eta_i$ axis. We write
\begin{equation}\label{Ddef}
\mathbf D_{i,j}=D e_{i,j}\hat y_i
\end{equation}
where $e_{i,j}$ =+1 (-1) if $j>i$ ($j<i)$ for NN $j$ on the $\hat x$ or $\hat z$ axis.
One has by definition $e_{j,i}=-e_{i,j}$.

\begin{figure}[ht!]
\centering
\includegraphics[width=8cm,angle=0]{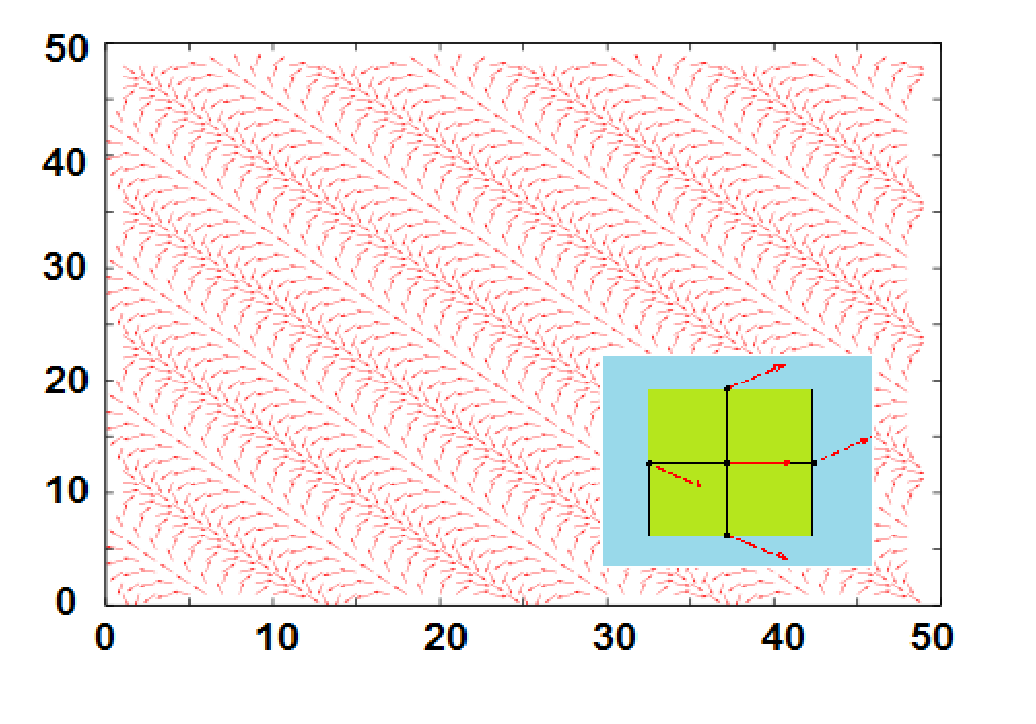}
\caption{The ground state is a planar configuration on the $xz$ plane. The figure shows the case where
$\theta=\pi/6$ ($D=-0.577)$, $J_1=J_\bot =1$ using
the steepest descent method. The inset shows a zoom around a spin with its nearest neighbors.\label{ffig1}}
\end{figure}

The easiest way to determine the GS is to minimize the local energy at each spin:  taking a spin and calculating  the local field
acting on it from its neighbors. Then, we align the spin in its local-field direction to minimize its energy.
Repeating this procedure for all spins, we say we realize one sweep.  We have to make a sufficient number of sweeps  to obtain the convergence  with a
desired precision (see details in Ref. \cite{NgoSurface}).  This local energy minimization is called "the steepest descent method".
We show  in Fig. \ref{ffig1} the configuration obtained
for $D=-0.5$ using $J_1=J_2=1$.

We see that each spin has the same angle with its four NN in the plane
(angle between NN in adjacent planes is zero). We demonstrate now the dependence of $\theta$ on $J_1$:
the energy of the spin $\mathbf S_i$ is written as
\begin{equation}
E_{i}=-4J_1S^2\cos\theta-2J_2 S^2+ 4DS^2\sin\theta
\end{equation}
where $\theta=|\theta_{i,j}|$
minimizing $E_i$ with respect to $\theta$ one obtains
\begin{equation}
\frac{dE_{i}}{d\theta}=0\  \ \Rightarrow \  \  -\frac{D}{J_1}=\tan \theta
\  \ \Rightarrow \  \  \theta=\arctan (-\frac{D}{J_1})\label{gsangle}
\end{equation}
The result is in agreement with that obtained by
the steepest descent method. An example has been shown in Fig. \ref{ffig1}.

We rewrite the DM term of Eq. (\ref{eqn:hamil3})  as
\begin{eqnarray}
\mathbf S_i\times\mathbf S_j&=& (-S_i^zS_j^y-S_i^yS_j^x\sin \theta_{i,j}+S_i^yS_j^z\cos \theta_{i,j})\hat \xi_i\nonumber\\
&&+(S_i^xS_j^x\sin\theta_{i,j}+S_i^zS_j^z\sin\theta_{i,j})\hat \eta_i\nonumber\\
&&+(S_i^xS_j^y-S_i^yS_j^z\sin \theta_{i,j}-S_i^yS_j^x\cos \theta_{i,j})\hat \zeta_i
\end{eqnarray}
From Eq. (\ref{Ddef}), we obtain
\begin{eqnarray}
\mathcal H_{DM}&=&\sum_{\left<i,j\right>}\mathbf D_{i,j}\cdot \mathbf S_i\times\mathbf S_j\nonumber\\
&=&D\sum_{\left<i,j\right>}(S_i^xS_j^xe_{i,j}\sin\theta_{i,j}+S_i^zS_j^ze_{i,j}\sin\theta_{i,j})\nonumber\\
&=&\frac{D}{4}\sum_{\left<i,j\right>}[(S_i^++S_i^-)(S_j^++S_j^-)e_{i,j}\sin\theta_{i,j}\nonumber\\
&&+4S_i^zS_j^ze_{i,j}\sin\theta_{i,j}]\nonumber\\
&&\label{DMterm}
\end{eqnarray}
where we have replaced $S^x$ by $(S^++S^-)/2$.
Note that $e_{i,j}\sin\theta_{i,j}$ is always positive since for a NN on the positive axis direction,
$e_{i,j}=1$ and
$\sin\theta_{i,j}=\sin \theta$ where $\theta$ is positively defined, while for a NN on the negative axis direction,
$e_{i,j}=-1$ and $\sin\theta_{i,j}=\sin(-\theta)=-\sin\theta$.

\subsection{Formulation of the Green's function technique for the Dzyaloshinskii-Moriya system}\label{DMIGF}

Using the transformation into the local coordinates, Eqs. (\ref{SI})-(\ref{local6}), one has
\begin{eqnarray}
\mathcal H &=& - \sum_{<i,j>}
J_{i,j}\Bigg\{\frac{1}{4}\left(\cos\theta_{i,j} -1\right)
\left(S^+_iS^+_j +S^-_iS^-_j\right)\nonumber\\
&+& \frac{1}{4}\left(\cos\theta_{i,j} +1\right) \left(S^+_iS^-_j
+S^-_iS^+_j\right)\nonumber\\
&+&\frac{1}{2}\sin\theta_{i,j}\left(S^+_i +S^-_i\right)S^z_j
-\frac{1}{2}\sin\theta_{i,j}S^z_i\left(S^+_j
+S^-_j\right)\nonumber\\
&+&\cos\theta_{i,j}S^z_iS^z_j\Bigg\}\nonumber\\
&+&\frac{D}{4}\sum_{\left<i,j\right>}[(S_i^++S_i^-)(S_j^++S_j^-)e_{i,j}\sin\theta_{i,j}\nonumber\\
&&+4S_i^zS_j^ze_{i,j}\sin\theta_{i,j}]\nonumber\\
&&\label{HGH2}
\end{eqnarray}
Note that the quantization axes of the spins are in the $xz$ planes as shown in Fig. \ref{localco}.

We emphasize that while the sinus term of the DM Hamiltonian, Eq. (\ref{DMterm}), remain after summing over the NN,  the sinus terms of $\mathcal H_e$,
the 3rd line of Eq. (\ref{HGH2}), are zero after summing over opposite NN because there is no $e_{i,j}$ term.

It is very important to emphasize again that the commutation relations between spin operators
$S^z$ and $S^{\pm}$ are valid when the spin lies on its local quantization axis.  Therefore, it is necessary ro use the local coordinates for each spin.

In two dimensions (2D) there is no long-range order at non-zero $T$ for isotropic
spin models with short-range interaction \cite{Mermin}.
Thin films have very small thickness, not far from 2D systems.  Thus, in order to stabilize the ordering at very low $T$, we use a very small anisotropy interaction between between $\mathbf S_i$
and $\mathbf S_j$ as follows
\begin{equation}
\mathcal H_a= -\sum_{<i,j>} I_{i,j}S^z_iS^z_j\cos\theta_{i,j}
\end{equation}
where $I_{i,j}(>0)$ is positive, small compared to $J_1$, and limited to NN in the $xz$ plane.
For simplicity, we suppose $I_{i,j}=I_1$ for all such NN pairs. As we will see below,  the small value of $I_1$ does stabilize the SW spectrum when $D$ becomes large.  The Hamiltonian is finally given by
\begin{equation}\label{totalH}
\mathcal H=\mathcal H_e+\mathcal H_{DM}+\mathcal H_a
\end{equation}


Using the two GF's in the real space given by Eqs. (\ref{green59a})-(\ref{green60}) and
using the same method, we study the effect of the DM interaction.
For the DM term, the commutation relations $[\mathcal H,S_i^{\pm}]$ lead to:
\begin{equation}
D\sum_l\sin \theta [\mp S_i^z(S_l^{+}+S_l^{-})+\pm 2S_i^{\pm} S_l^z]\label{HDMC}
\end{equation}
which gives rise, using the Tyablikov decoupling,  to the following GF's:
\begin{equation}\label{DMa}
<<S_i^zS_l^{\pm};S_j^->>\simeq <S_i^z><<S_l^{\pm};S_j^->>
\end{equation}
These functions are in fact the  $G$ and $F$ functions. There are thus no new GF's generated by the equations of motion.

As in section \ref{heli}, the Fourier
transforms in the $xz$ plane $ g_{n,n'}$ and $  f_{n,n'}$ of the $G$ and $F$  lead  to the matrix equation
\begin{equation}
\mathbf M \left( E \right) \mathbf h = \mathbf u,
\end{equation}
$\mathbf M\left(E\right)$ being given by Eq. (\ref{eq:HGMatrixM}) below
\begin{equation}\label{eq:HGMatrixM}
\left(%
\begin{array}{ccccccccc}
  E+A_1& B_1    & C_1& 0&0& 0& 0&0&0\\
   -B_1   & E-A_1  & 0 & -C_1 &0&0&0&0&0\\
   \cdots & \cdots & \cdots &\cdots&\cdots&\cdots&\cdots&\cdots&\cdots\\
 \cdots&0&C_{n}&0&E+A_{n}&B_n&C_{n}&0&0\\
 \cdots&0&0&-C_{n}&-B_n&E-A_{n}&0&-C_{n}&0\\
         \cdots  & \cdots & \cdots &\cdots&\cdots&\cdots&\cdots&\cdots&\cdots\\
  0& 0&0& 0& 0  & C_{N}   & 0   &E + A_{N}&B_{N}\\
  0&0&0&0& 0 & 0  & -C_{N}& -B_{N}& E-A_{N}\\
\end{array}%
\right)
\end{equation}
where $E=\hbar \omega$ is the SW energy and the matrix elements are given by
\begin{eqnarray}
A_{n} &=& -J_1[8<S^z_n>\cos\theta (1+d_n)\nonumber\\
&&- 4 <S^z_n>\gamma (\cos\theta+1)]\nonumber\\
&&-2J_2 (<S^z_{n-1}>+<S^z_{n+1}>)\nonumber\\
&&-8D\sin \theta < S^z_{n}>\gamma\nonumber\\
&&+8D\sin \theta < S^z_{n}>\label{anterm}\\
B_n &=& 4J_1 < S^z_{n}> \gamma (\cos\theta-1)\nonumber\\
&&-8D \sin \theta < S^z_{n}>\gamma\label{bnterm}\\
C_n &=& 2J_2 < S^z_{n}> \label{cnterm}
\end{eqnarray}
where $n=1,2,...,N$ denoting the layer numbers, $d_n=I_1/J_1$, $\gamma=(\cos k_xa+\cos k_za)/2$,
$k_{x}$ and  $k_{z}$ are the wave-vector components in the $xz$ planes, $a$ being the lattice constant.
Remarks: (i) if $n=1$ (surface layer) then there are no  $n-1$ terms in the $A_n$, (ii) if $n=N$
then there are no  $n+1$ terms in $A_n$.

For a thin film, the SW frequecies at a given wave vector $\vec k=(k_x,kz)$
are obtained by diagonalizing  (\ref{eq:HGMatrixM}).

The magnetization of the layer $n$ at finite $T$ is calculated as in the helimagnetic case shown in the previous section.
The formula of the zero-point spin contraction   is also presented there.
The transition temperature $T_c$ can be also calculated by the same method. Let us show in the following the results.

\subsection{Results for 2D and 3D cases}
In the 2D case, one has only one layer. The matrix (\ref{eq:HGMatrixM}) is
\begin{eqnarray}
(E+A_n)g_{n,n'}+B_nf_{n,n'}&=&2< S^z_{n}>\delta(n,n')\nonumber\\
-B_ng_{n,n'}+(E-A_n)f_{n,n'}&=&0
\end{eqnarray}
where $A_n$ is given by (\ref{anterm}) but without $J_2$ term for the 2D case.
Coefficients $B_n$
is given by (\ref{bnterm}) and  $C_n=0$.
The SW frequencies are determined by the following secular equation
\begin{eqnarray}
&&(E+A_n)(E-A_n)+B_n^2=0\nonumber\\
&\rightarrow&E^2-A_n^2+B_n^2=0\nonumber\\
&\rightarrow&E=\pm \sqrt{(A_n+B_n)(A_n-B_n)}\label{SWE}
\end{eqnarray}
Several remarks are in order:

(i) when $\theta=0$,  the last three terms of $A_n$ and $B_n$  are zero: one recovers the ferromagnetic SW dispersion relation

\begin{equation}
E=2ZJ_1<S_n^z>(1-\gamma)
\end{equation}
where $Z=4$ is the coordination number of the
square lattice (taking $d_n=0$),

(ii)  when $\theta=\pi$, one has $A_n=8J_1<S_n^z>$, $B_n=-8J_1<S_n^z>\gamma$.
One recovers then the antiferromagnetic SW dispersion relation
\begin{equation}
E=2ZJ_1<S_n^z>\sqrt{1-\gamma^2}
\end{equation}

\begin{figure}[ht!]
\centering
\includegraphics[width=8cm,angle=0]{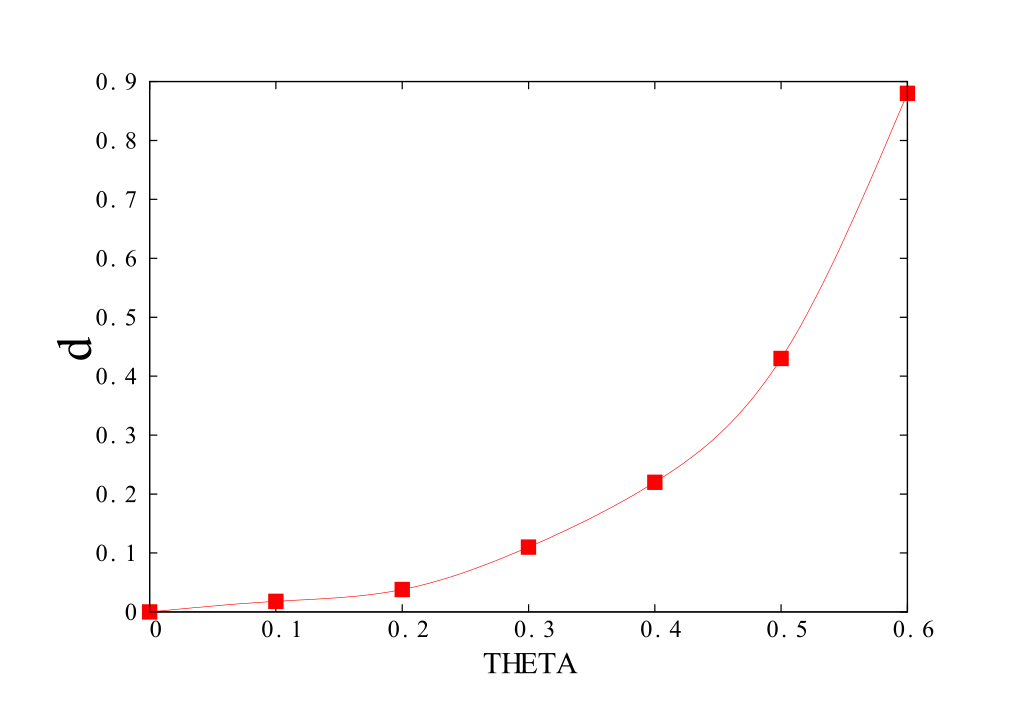}
\caption{Value $d_c$ at which $E=0$ at $\gamma=1$ ($\vec k=0)$ vs $\theta$ (in radian).
Above this value, $E$ is real. See text for
comments. \label{ffig3}}
\end{figure}

(iii) when there is a DM interaction, one has $0<\cos \theta < 1$ ($0<\theta<\pi/2$). If $d_n=0$, the
quantity in the square root of Eq. (\ref{SWE}) becomes negative at $\gamma=1$ when $\theta$ is not zero.
The SW spectrum is not stable at $k_x=k_y=0$ because the energy is not real. The anisotropy $d_n$ can remove
this instability if it is larger than a threshold value $d_c$. We solve the equation $(A_n+B_n)(A_n-B_n)=0$ to find $d_c$. In Fig. \ref{ffig3} we show  $d_c$ versus $\theta$. As seen, $d_c$ increases from zero with increasing $\theta$.

As we have anticipated, we need  to include an anisotropy in order to allow for SW to be excited even at $T=0$ and
for a long-range ordering at non-zero $T$ in 2D as seen below.

We show in Fig. \ref{ffig4}  the SW dispersion relation calculated from Eq. (\ref{SWE}) for $\theta=0.2$ and 0.6 (radian). As seen, the spectrum is symmetric for positive and negative wave vectors. It is also symmetric for left
and right precessions. One observes that for small $\theta$, namely small $D$, $E(k)$ is proportional to $k^2$ at low $k$ (see Fig. \ref{ffig4}a). This behavior is that in ferromagnets. For large $\theta$, one observes that $E(k)$ becomes linear in $k$ as seen in Fig. \ref{ffig4}b. This behavior is similar to that of antiferromagnets.  Note that the change of behavior is progressive with increasing $\theta$, we do not observe a sudden transition from $k^2$ to $k$ behavior.  This behavior is also observed in 3D and in thin films as well.

\begin{figure}[ht!]
\centering
\includegraphics[width=7cm,angle=0]{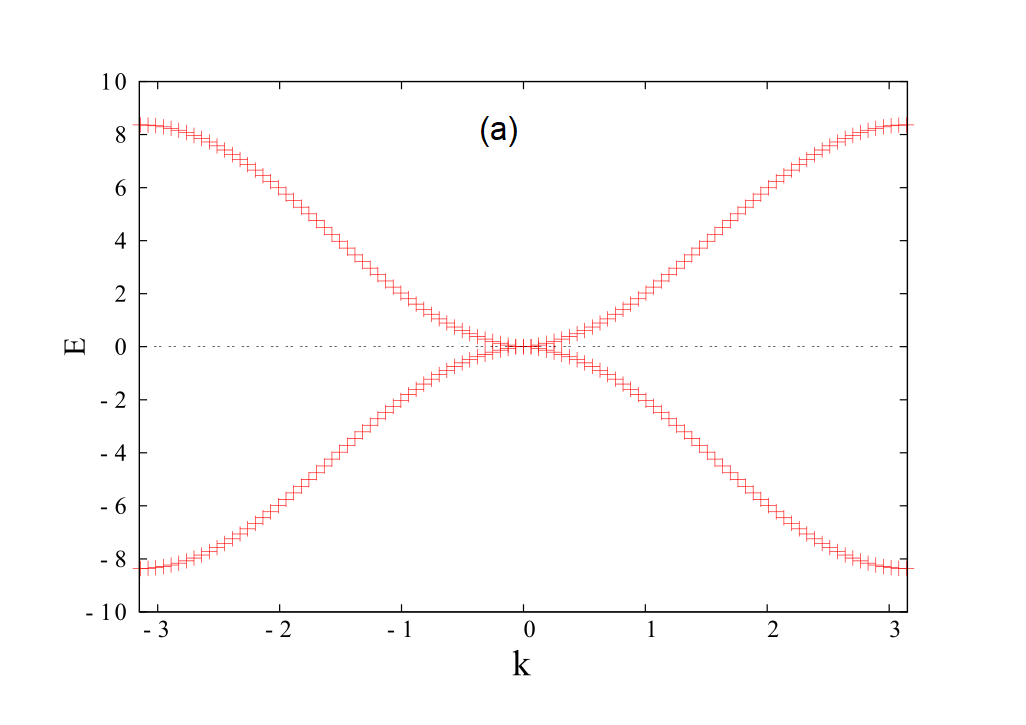}
\includegraphics[width=7cm,angle=0]{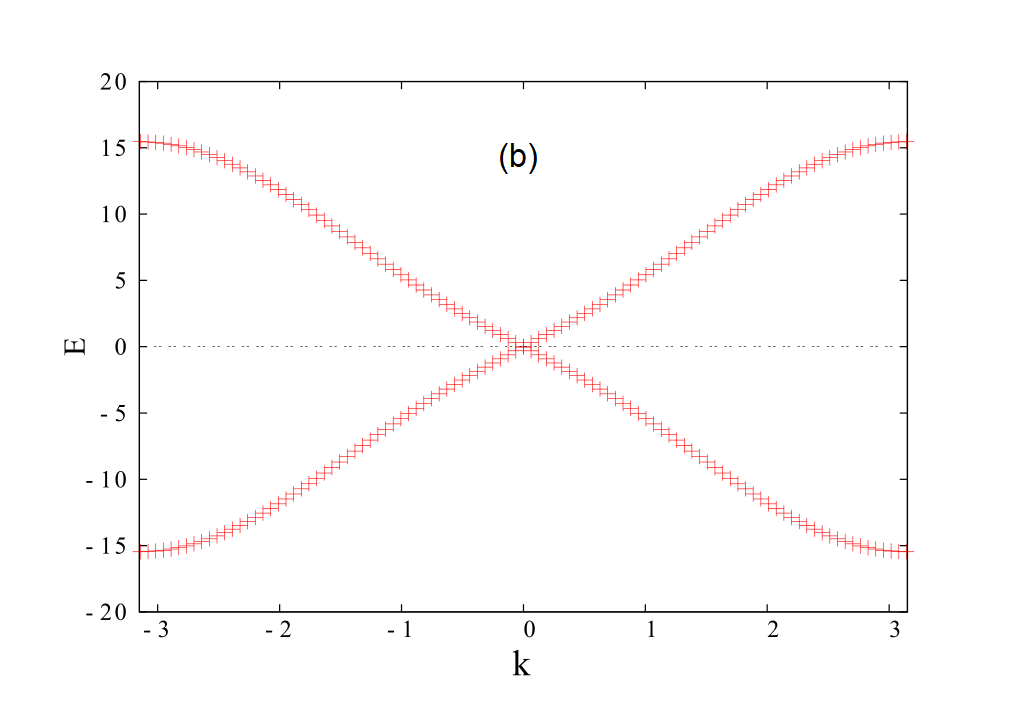}
\caption{SW frequency $E (k)$ as a function of $k\equiv k_x=k_z$ in the case (a) $\theta=0.2$ and (b) $\theta=0.6$ in 2D.  See text for detailed comments.
\label{ffig4}}
\end{figure}

\begin{figure}[ht!]
\centering
\includegraphics[width=7cm,angle=0]{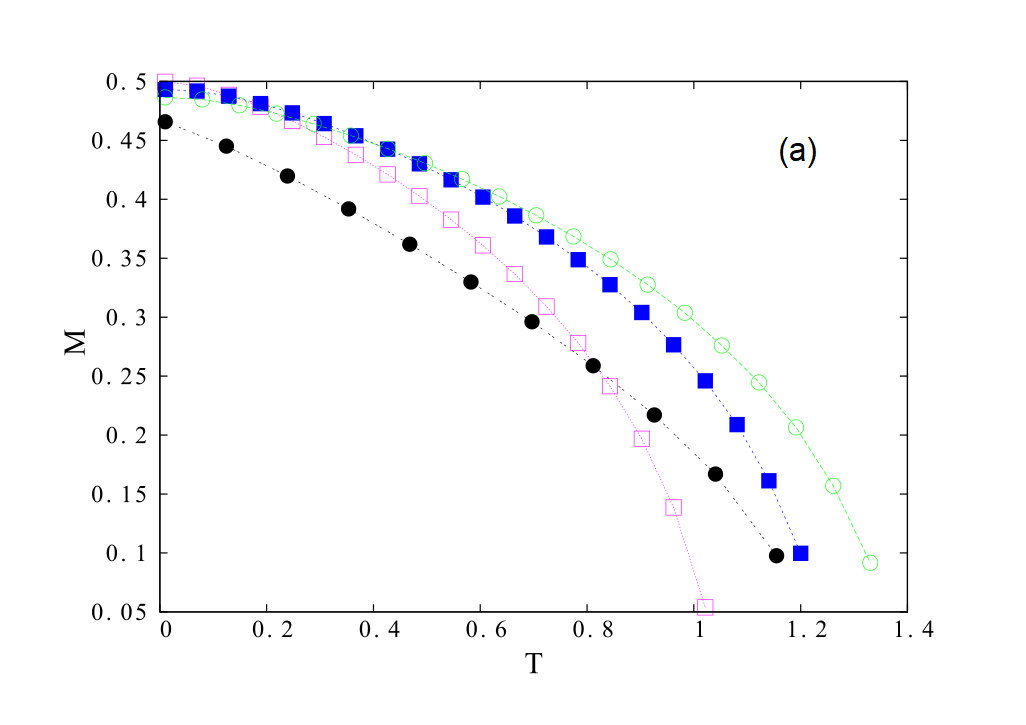}
\includegraphics[width=7cm,angle=0]{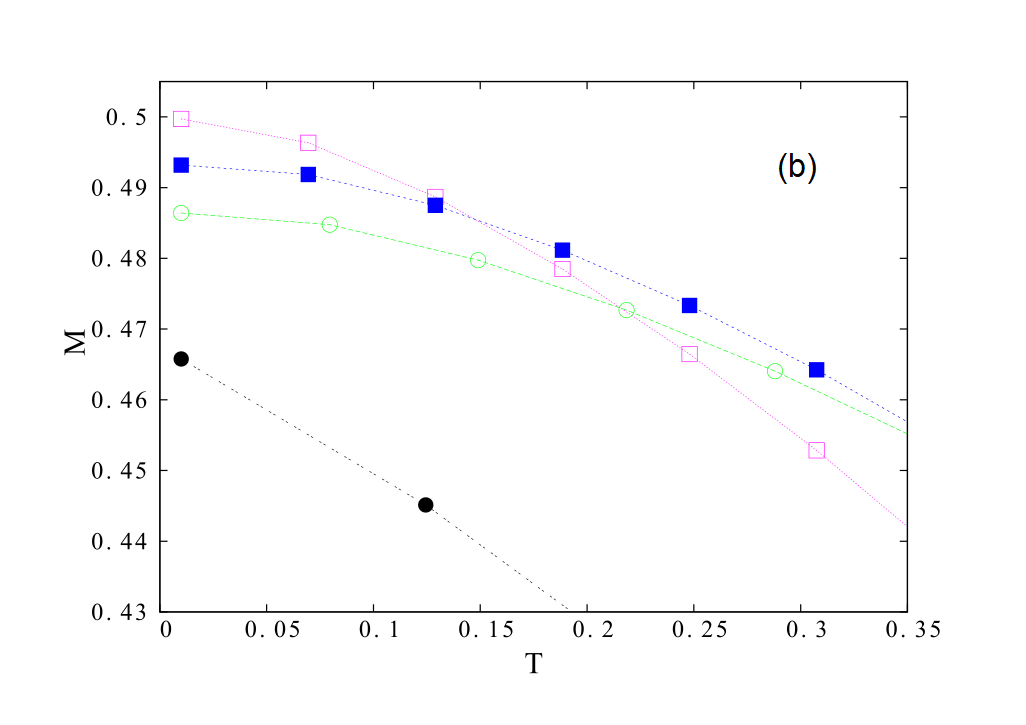}
\caption{(a) Magnetization $M$ as a function of  $T$ for the 2D case with $\theta=0.1$, $\theta=0.3$, $\theta=0.4$,  $\theta=0.6$ (void magenta squares, blue filled squares, green void circles
and filled black circles, respectively), (b) Cross-over of magnetizations is enlarged at low $T$. See text for comments.
\label{ffig5}}
\end{figure}

\begin{figure}[ht!]
\centering
\includegraphics[width=8cm,angle=0]{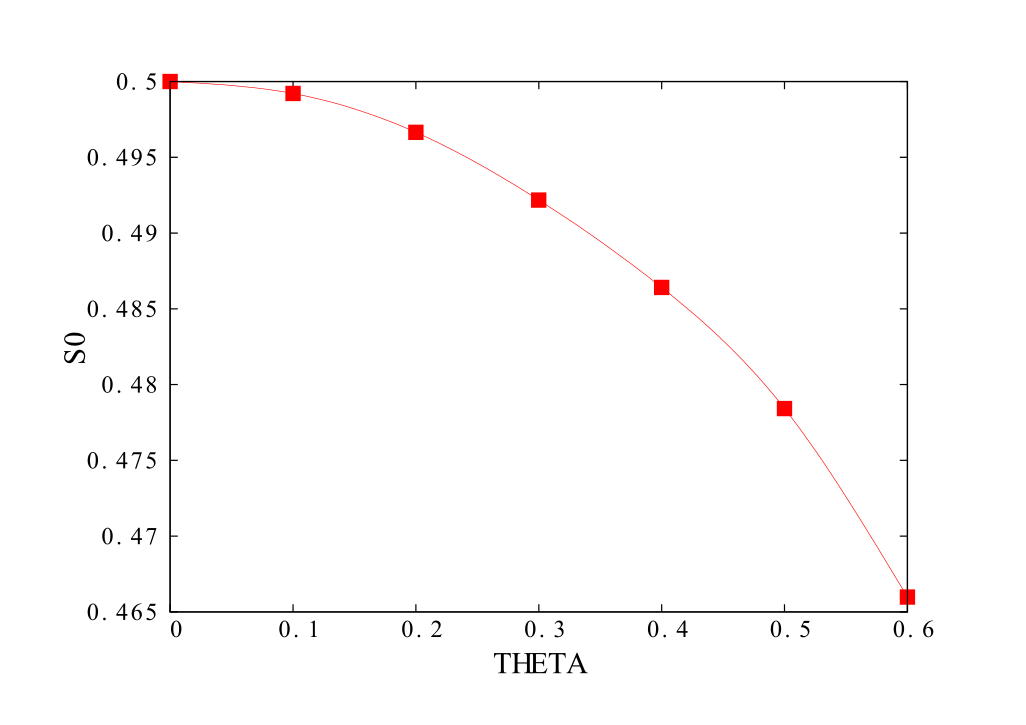}
\caption{Spin length at $T=0$ for the 2D case as a function of $\theta$ (radian).
\label{ffig6}}
\end{figure}

As said earlier, the inclusion of an anisotropy $d$ permits to have a llong-range ordering at  $T\neq 0$
in 2D:  Fig. \ref{ffig5} displays the magnetization $M$ ($\equiv <S^z>$) calculated by Eq. (\ref{lm2H}) where in each case the
limit value $d_c$ has been  used. We note that $M$ depends strongly on $\theta$: at high $T$ the larger $\theta$ the stronger $M$. However, at $T=0$ the spin length is smaller for larger $\theta$ due to the zero-point spin contraction \cite{DiepTM} calculated by Eq. (\ref{surf38}).  As a consequence there is a cross-over of layer magnetizations at low $T$ as shown in Fig. \ref{ffig5}b.  The spin length at $T=0$ is shown in Fig. \ref{ffig6} for several $\theta$.

%

We now consider the 3D case. The crystal is infinite in three direction. The Fourier transform in the $y$ direction, namely $g_{n\pm 1}=g_n\mbox{e}^{\pm ik_ya}$ and $f_{n\pm 1}=f_n\mbox{e}^{\pm ik_ya}$ reduces the matrix  (\ref{eq:HGMatrixguH})  to two coupled equations of $g$ and $f$ functions. One has
\begin{eqnarray}
(E+A')g + Bf&=&2<S^z>\nonumber\\
-Bg+(E-A')f&=&0
\end{eqnarray}
where
\begin{eqnarray}
A' &=& -J_1[8<S^z>\cos\theta (1+d)\nonumber\\
&&- 4 <S^z>\gamma (\cos\theta+1)]\nonumber\\
&&-4J_2 <S^z>\nonumber\\
&&+4J_2 < S^z>\cos(k_ya)\nonumber\\
&&-8D\sin \theta < S^z>\gamma\nonumber\\
&&+8D\sin \theta < S^z>\label{anterm3D}\\
B &=& 4J_1 < S^z> \gamma (\cos\theta-1)\nonumber\\
&&-8D \sin \theta < S^z>\gamma\label{bnterm3D}
\end{eqnarray}
The spectrum is given by
\begin{equation}\label{3Dspect}
E=\pm\sqrt{(A'+B)(A'-B)}
\end{equation}
In the ferromagnetic case, $\cos \theta=1$, thus $B=0$. Arranging the Fourier transforms in three directions,   one gets the 3D ferromagnetic dispersion relation $E=2Z<S^z>(1-\gamma^2)$ where $\gamma=[\cos (k_xa)+\cos (k_ya)+\cos (k_za)]/3$
and $Z=6$, coordination number of the simple cubic lattice.

\begin{figure}[ht!]
\centering
\includegraphics[width=8cm,angle=0]{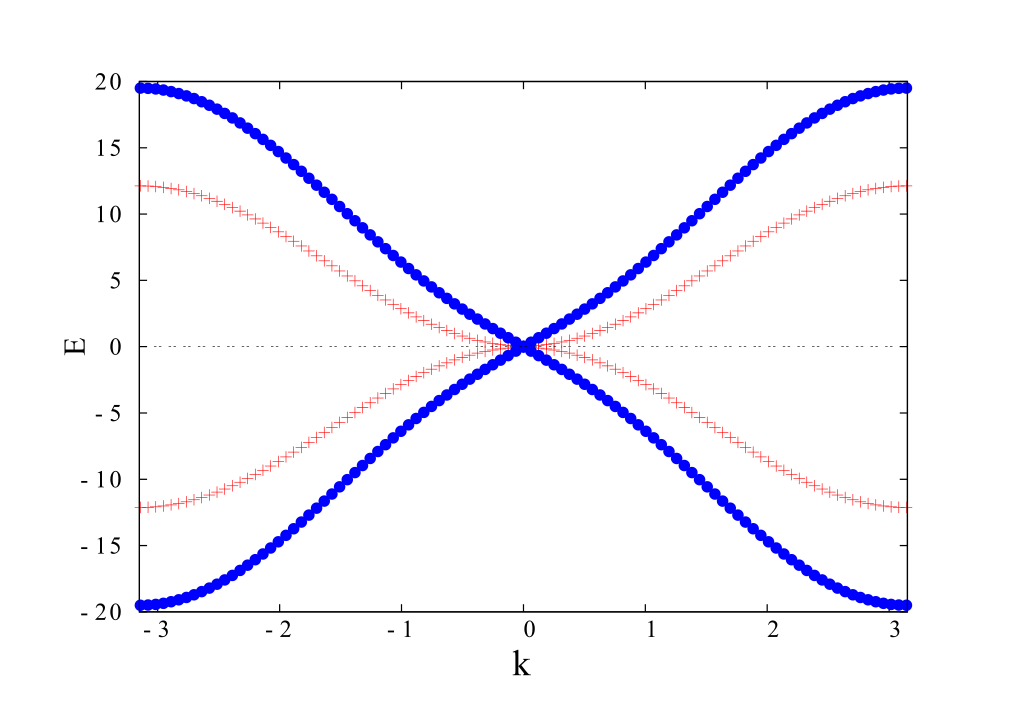}
\caption{Spin-wave spectrum $E (k)$ versus $k\equiv k_x=k_z$ for  $\theta=0.1$ (red crosses) and  $\theta=0.6$ (blue circles) in three dimensions. Note the linear-$k$ behavior at low $k$ for the large value of $\theta$. See text for comments.
\label{ffig7}}
\end{figure}

As in the 2D case, we find a threshold value $d_c$ for which is the same for a given $\theta$. This is rather obvious because the DM interaction operates in the plane making an angle $\theta$ between spins
in the plane, therefore its effects act on SW in each plane, not in the $y$ direction perpendicular to the "DM planes".  Using Eq. (\ref{3Dspect}), we calculate the 3D spectrum displayed in Fig. \ref{ffig7} for a small and a large value of $\theta$. As in the 2D case, we observe $E\propto k$ when $k\rightarrow 0$ for large $\theta$.
The main properties of the system are thus governed by the in-plane DM interaction.

\begin{figure}[ht!]
\centering
\includegraphics[width=8 cm,angle=0]{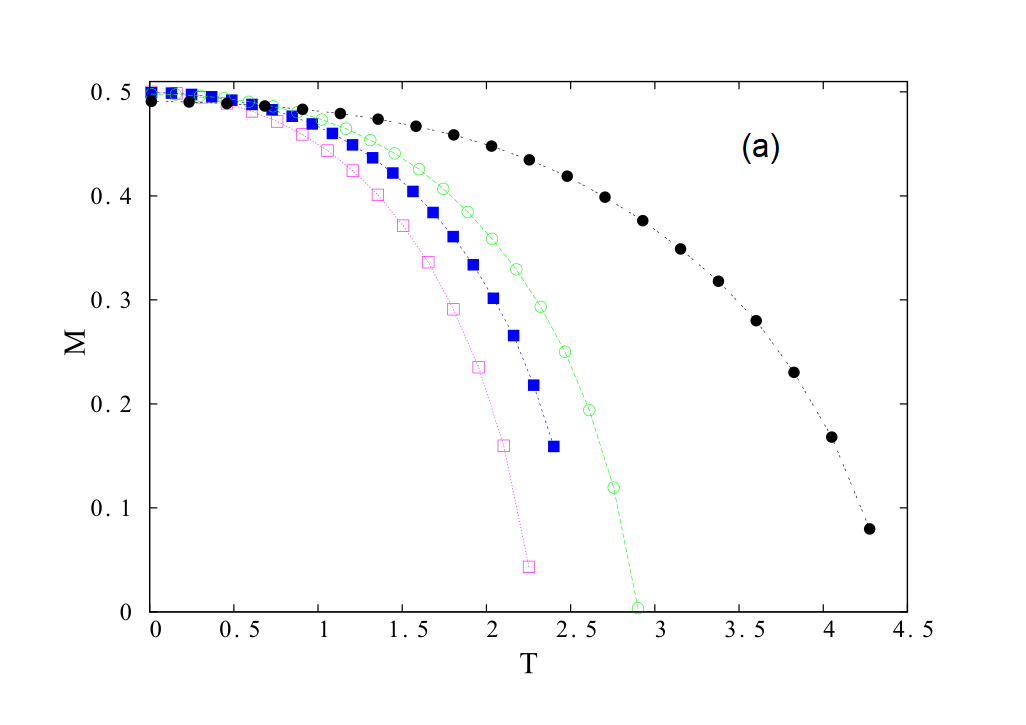}
\includegraphics[width=7cm,angle=0]{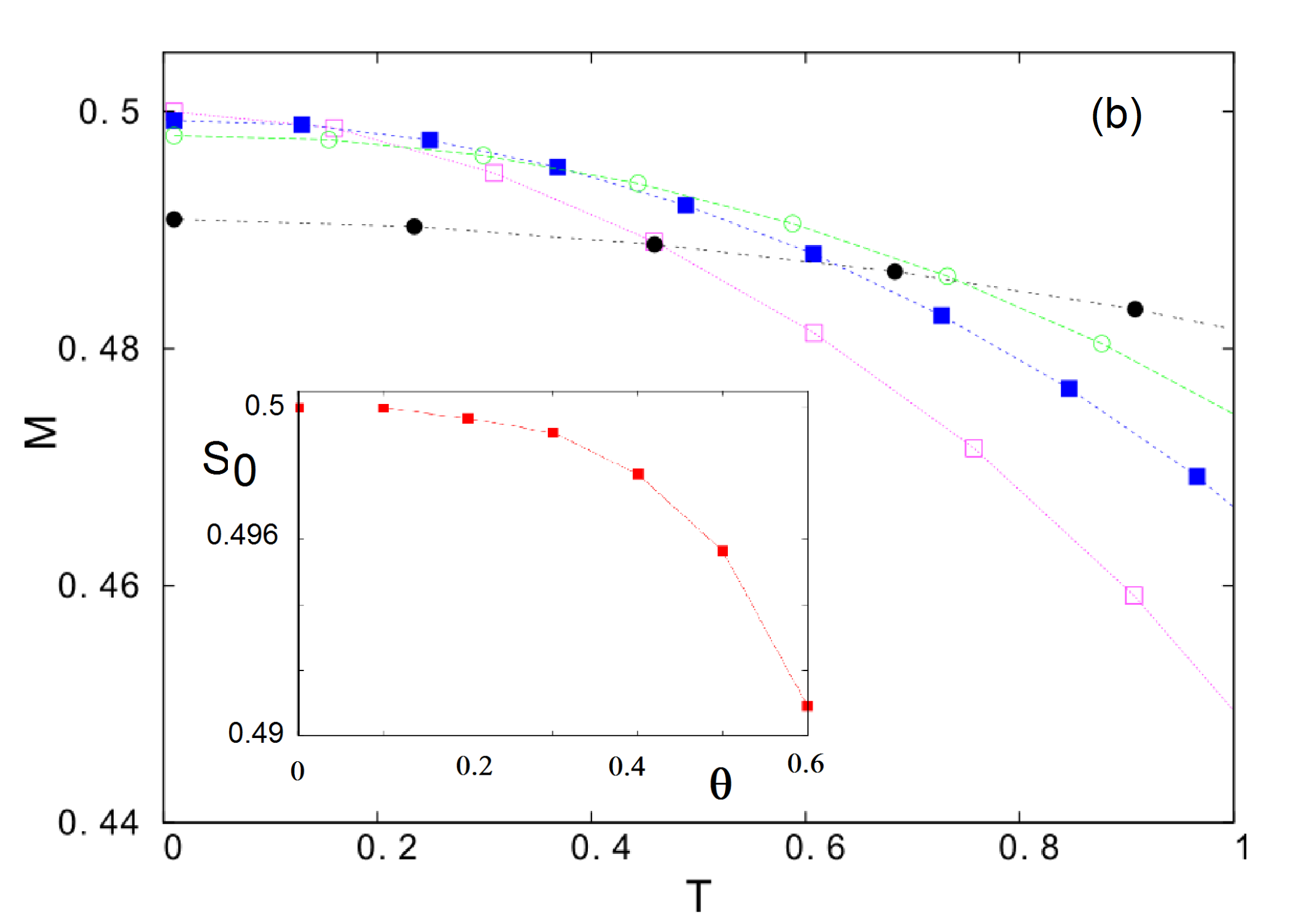}
\caption{(a) Magnetization $M$ versus temperature $T$ for a 3D crystal $\theta=0.1$ (radian), $\theta=0.3$,  $\theta=0.4$, $\theta=0.6$ (void magenta squares, blue filled squares, green void circles
and filled black circles, respectively), (b) Zoom to show the cross-over of magnetizations at low $T$ for different $\theta$, inset shows $S_0$ versus $\theta$. See text for comments.
\label{ffig8}}
\end{figure}

Figure \ref{ffig8} displays the magnetization $M$ versus $T$ for several values of $\theta$.  As in the 2D case, when the DM interaction is included, the spins undergo a zero-point contraction which increases with increasing $\theta$. The competition between quantum fluctuations at $T=0$ and thermal effects at high $T$ gives rise to magnetization cross-over shown in Fig. \ref{ffig8}b. The spin length at $T=0$ vs $\theta $ is shown in the inset of Fig. \ref{ffig8}b. Comparing these results to those of the 2D case, we see that the spin contraction in 2D is stronger than in 3D. This is physically expected because quantum fluctuations are stronger at lower dimensions.

\subsection{The case of a thin film}\label{results}

As in 2D and 3D cases, in the case of a thin film it is necessary to use a value for $d_n$
larger or equal to $d_c$ given in Fig. \ref{ffig3} to stabilize the SW at long wave-length.
Note that for thin films with more than one layer, the value of $d_c$
calculated for the 2D case remains valid.

\begin{figure}[ht!]
\centering
\includegraphics[width=7cm,angle=0]{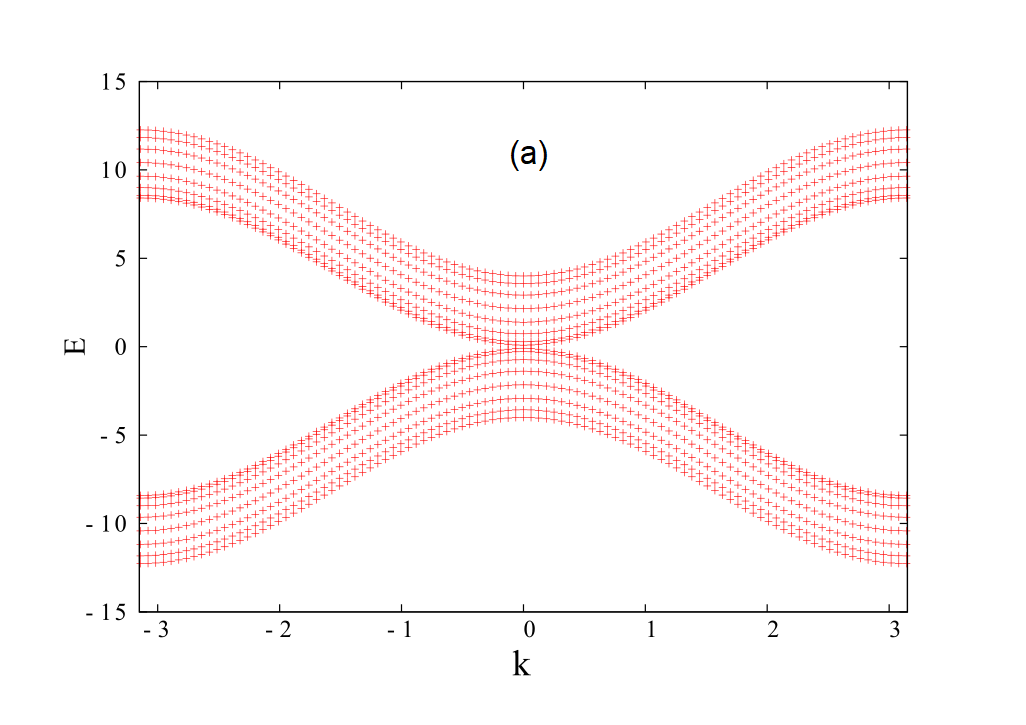}
\includegraphics[width=7cm,angle=0]{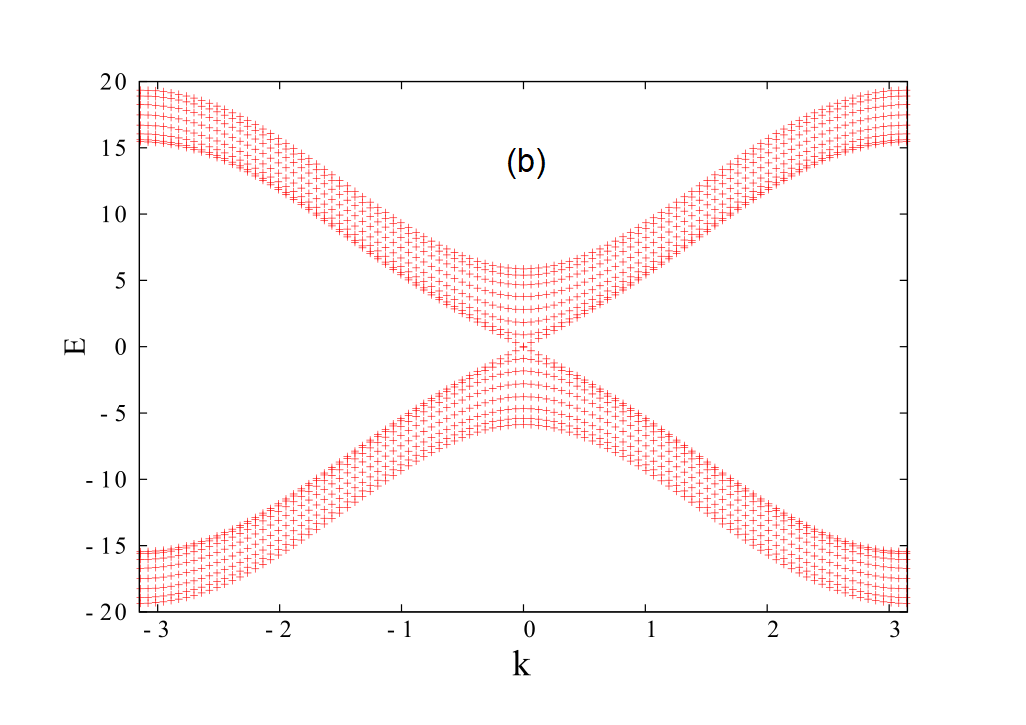}
\caption{Spin-wave spectrum $E (k)$ versus $k\equiv k_x=k_z$ for a thin
film of 8 layers: (a) $\theta=0.2$ (in radian) (b) $\theta=0.6$, using $d_c$ for each case.
Positive and negative branches correspond to right and left precessions.
Note the linear-$k$ behavior at low $k$. See text for comments.
\label{ffig9}}
\end{figure}

 Figure \ref{ffig9} displays the SW spectrum of a film of 8 layers with $J_1=J_2=1$ for
a small and a large $\theta$. As in the previous cases,  $E$ is proportional to $k$ for large $\theta$
(cf. Fig. \ref{ffig9}b) but only for the first mode. The higher modes are proportional to $k^2$.

\begin{figure}[ht!]
\centering
\includegraphics[width=5cm,angle=0]{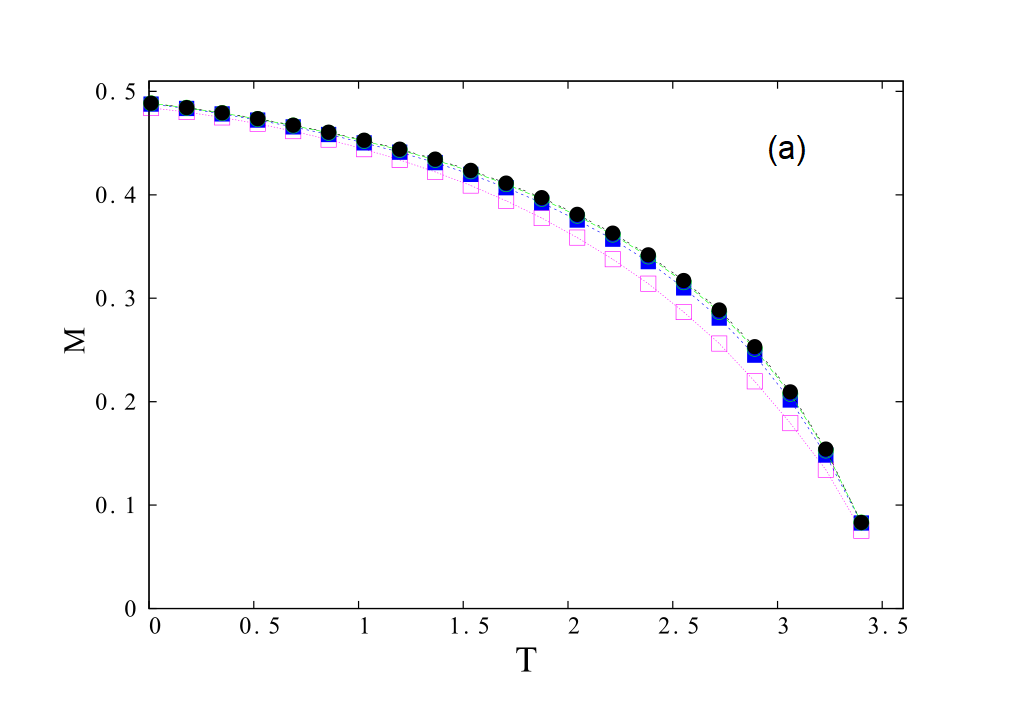}
\includegraphics[width=5cm,angle=0]{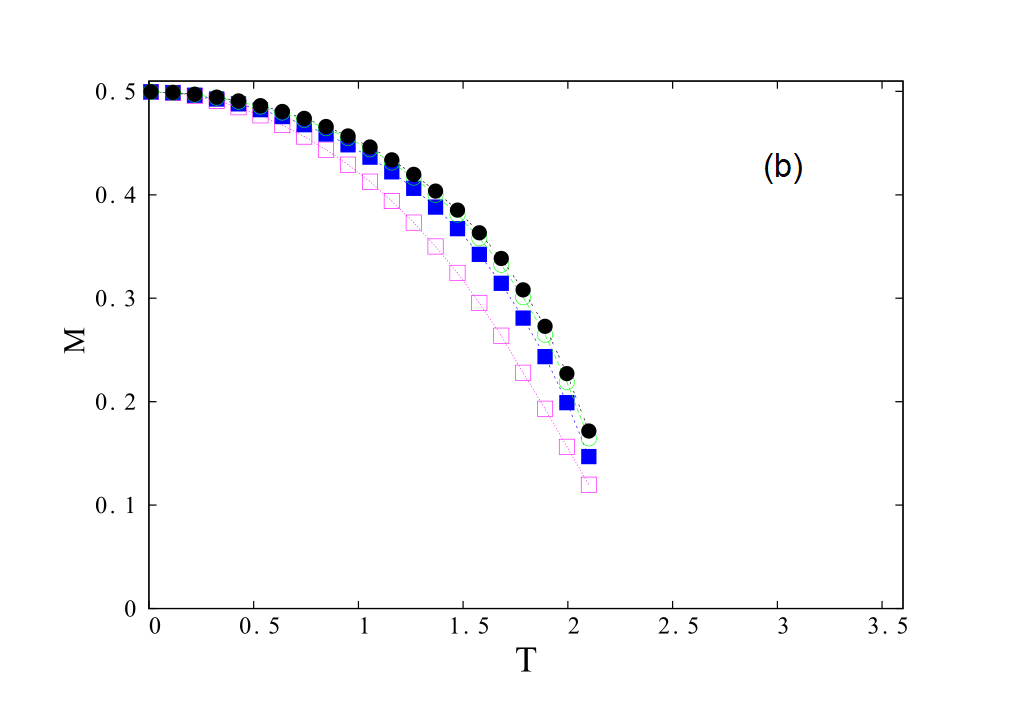}
\includegraphics[width=5cm,angle=0]{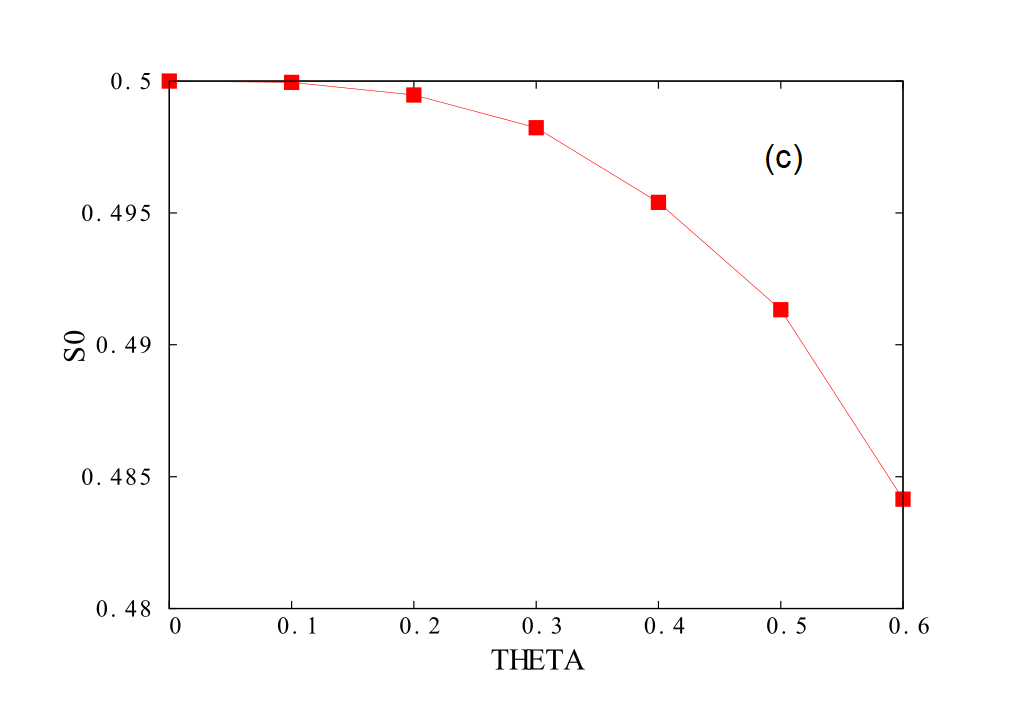}
\caption{Layer magnetizations $M$ versus temperature $T$ for a
film with $N=8$: (a) $\theta=0.6$ (radian), (b) $\theta=0.2$, (c) $S_0$ versus $\theta$.
\label{ffig10}}
\end{figure}

Figure \ref{ffig10} shows the layer magnetizations of the first four layers
in a 8-layer film (the other half is symmetric) for two values of $\theta$.
One observes that the surface magnetization is smaller than the magnetizations of other interior layers.
This is due to the lack of neighbors for surface spins \cite{DiepGF1979}.

The spin contraction  at $T=0$ is displayed  Fig. \ref{ffig10}c.

The effects of the surface exchange and the film thickness have been shown in Ref. \cite{Diep2017}.

To close this section, let us mention our work \cite{sharafullin2019dzyaloshinskii} on the DM interaction in magneto-ferroelectric superlattices where the SW in the magnetic layer have been calculated.  We have also studied the stability of skyrmions at finite $T$ in that work and in Refs. \cite{ElHog2018,Sharafullin2020}.

\section{Effect of Dzyaloshinskii-Moriya interaction in a frustrated antiferromagnetic triangular lattice}\label{DMAFTL}
The results of this section are not yet published \cite{Sahbi2022}.  We will not present this model in details.
We show the Hamiltonian, the GS and the SW spectrum.

\subsection{Model - Ground State}\label{Model}

\begin{figure}[h!]
\centering
\includegraphics[width=6cm]{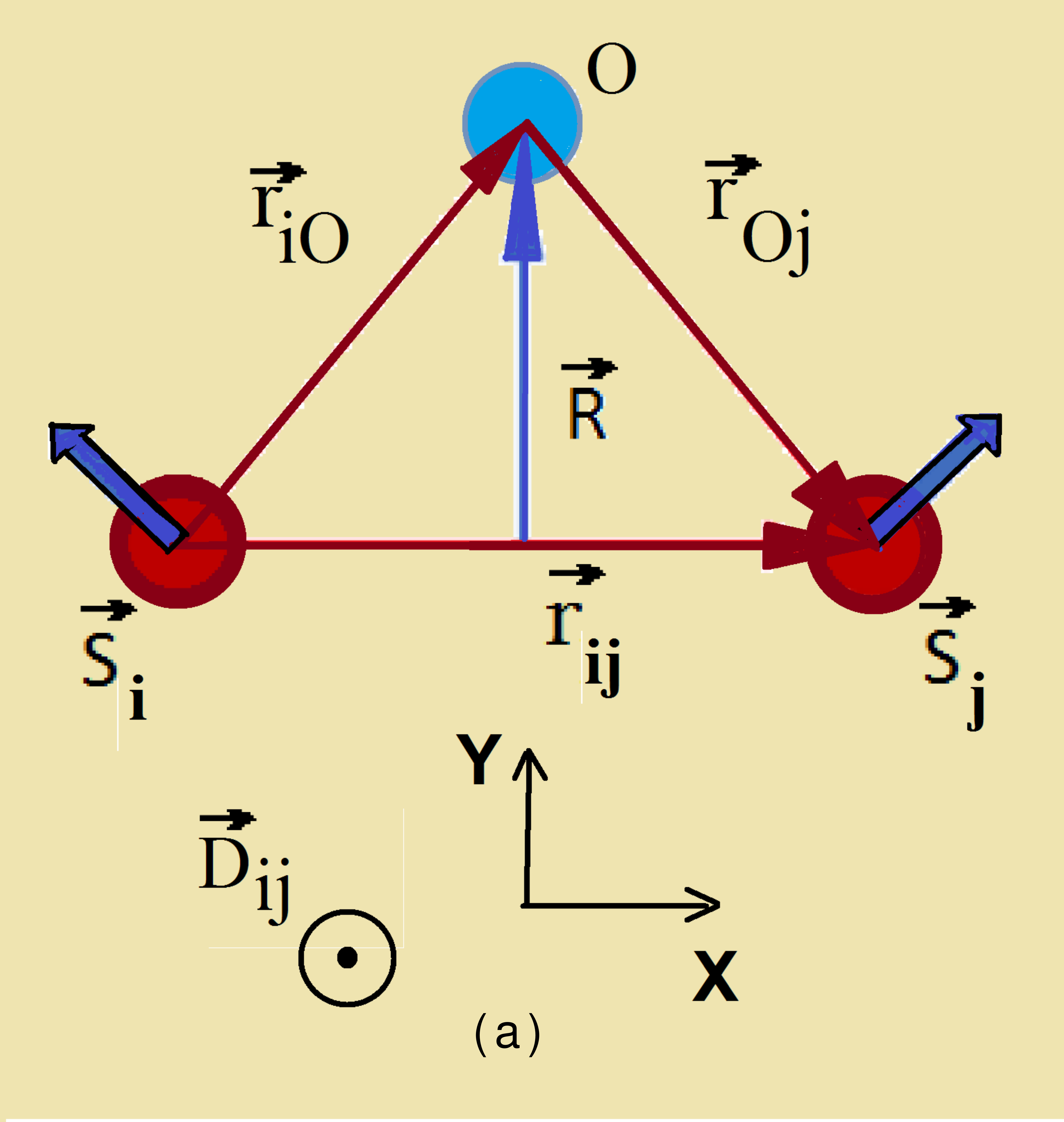}
\includegraphics[width=6cm]{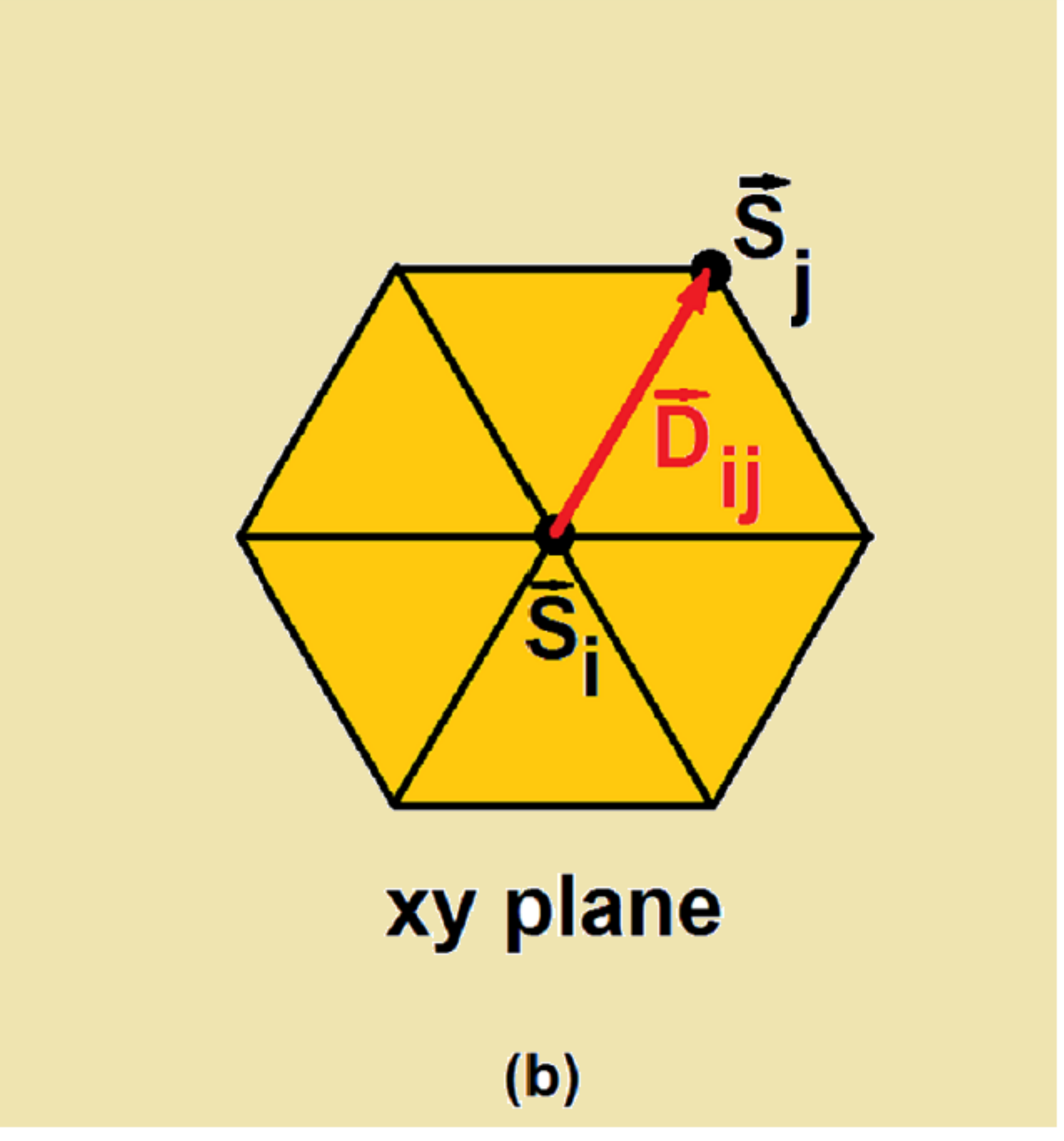}
\caption{(a)  $D$ vector along the $z$ direction perpendicular to the $xy$ plane. See the definition of the $D$ vector in the text, (b) In-plane $\mathbf D_{ij}$ vector chosen along the direction connecting spin $\mathbf S_i$ to spin $\mathbf S_j$ in the $xy$ plane.}\label{D}
\end{figure}

We consider a triangular lattice occupied by Heisenberg spins of magnitude 1/2. The DM interaction was introduced historically to explain the weak ferromagnetism in compounds MnO.  The superexchange between two Mn atoms is modified with the displacement of the oxygen atom between them.  If the displacement of the oxygen is in the $xy$ plane (see Fig. \ref{D}a), then the DM vector $\mathbf D_{i,j}$ is perpendicular to the $xy$ plane and is given by \cite{Keffer,Cheong}
\begin{equation}\label{D1}
\mathbf D_{i,j}\propto \mathbf r_{iO} \times \mathbf r_{Oj}\propto -\mathbf r_{ij} \times \mathbf R
\end{equation}
where $\mathbf r_{iO}=\mathbf r_O-\mathbf r_i$ and $\mathbf r_{Oj}=\mathbf r_j-\mathbf r_O$, $\mathbf r_{ij}=\mathbf r_j-\mathbf r_i$.   $\mathbf r_O$ is the position of non-magnetic ion (oxygen) and $\mathbf r_i$ the position of the spin $\mathbf S_i$ etc. These vectors are defined in Fig. \ref{D}a in the particular case where the displacements  are in the $xy$ plane. We have therefore $\mathbf D_{i,j}$ perpendicular to the $xy$ plane in this case.

Note however  that if the atom displacements are in 3D space, $\mathbf D_{i,j}$ can be in any direction. In this paper, we consider also the case where $\mathbf D_{i,j}$ lies in the $xy$ plane as shown in Fig. \ref{D}b where $\mathbf D_{i,j}$ is taken along the vector connecting spin $\mathbf S_i$ to spin $\mathbf S_{j}$.

Note that from Eq. (\ref{D1}) one has
\begin{equation}\label{D2}
\mathbf D_{j,i}=-\mathbf D_{i,j}
\end{equation}

In the case of perpendicular $\mathbf D_{i,j}$,
let us define $\mathbf u_{i,j}$ as the unit vector on the $z$ axis. From Eqs. (\ref{D1})-(\ref{D2}) one writes
\begin{eqnarray}
\mathbf D_{i,j}&=&D\mathbf u_{i,j}\label{D3a}\\
\mathbf D_{j,i}&=&D\mathbf u_{j,i}=-D\mathbf u_{i,j}\label{D3b}
\end{eqnarray}
where $D$ represents the DM interaction strength.  Note however that the DM interaction goes beyond the weak ferromagnetism and may find its origin in  various physical mechanisms. So, the form given in (\ref{D3a}) is a model, a hypothesis.

In the case of in-plane $\mathbf D_{i,j}$, we suppose that $\mathbf D_{i,j}$ is given as

\begin{equation}\label{PDM}
\mathbf D_{i,j}=D(\mathbf r_j-\mathbf r_i)/|\mathbf r_j-\mathbf r_i|=D\mathbf {r}_{ij}
\end{equation}
where $D$ is a constant and $\mathbf {r}_{ij}$ denotes the unit vector along $\mathbf r_j-\mathbf r_i$.
The case of in-plane $\mathbf D_{i,j}$ on the frustrated triangular lattice (see Fig. \ref{D}b) has been recently studied since this case gives rise to a beautiful skyrmion crystal composed of three interpenetrating sublattice skyrmions in a perpendicular applied magnetic field.\cite{Sahbi2022,Rosales,Mohylna}  A description of this case is however out of the purpose of this review.

\subsection{Ground State with a Perpendicular $\mathbf D$ in Zero Field}\label{GSH0}

The Hamiltonian is given by

\begin{eqnarray}
\mathcal{H}&=&-J \sum_{\langle ij \rangle} \mathbf {S_i} \cdot \mathbf{S_j} -
 D \sum_{\langle ij \rangle} \mathbf u_{i,j} \cdot \mathbf {S_i} \times \mathbf {S}_{j} \nonumber\\
&&-H \sum_i S_i^z
\end{eqnarray}
where $\mathbf {S_i}$ is a classical Heisenberg spin of magnitude 1 occupying the lattice site $i$. The first sum runs over all spin nearest-neighbor (NN) pairs  with an antiferromagnetic exchange interaction $J$ ($J<0$), while the second sum is performed over all  DM interactions between NN.  $H$ is the magnitude of a magnetic field applied along the $z$ direction perpendicular to the lattice $xy$ plane.

\begin{figure}[h!]
\centering
\includegraphics[width=12cm]{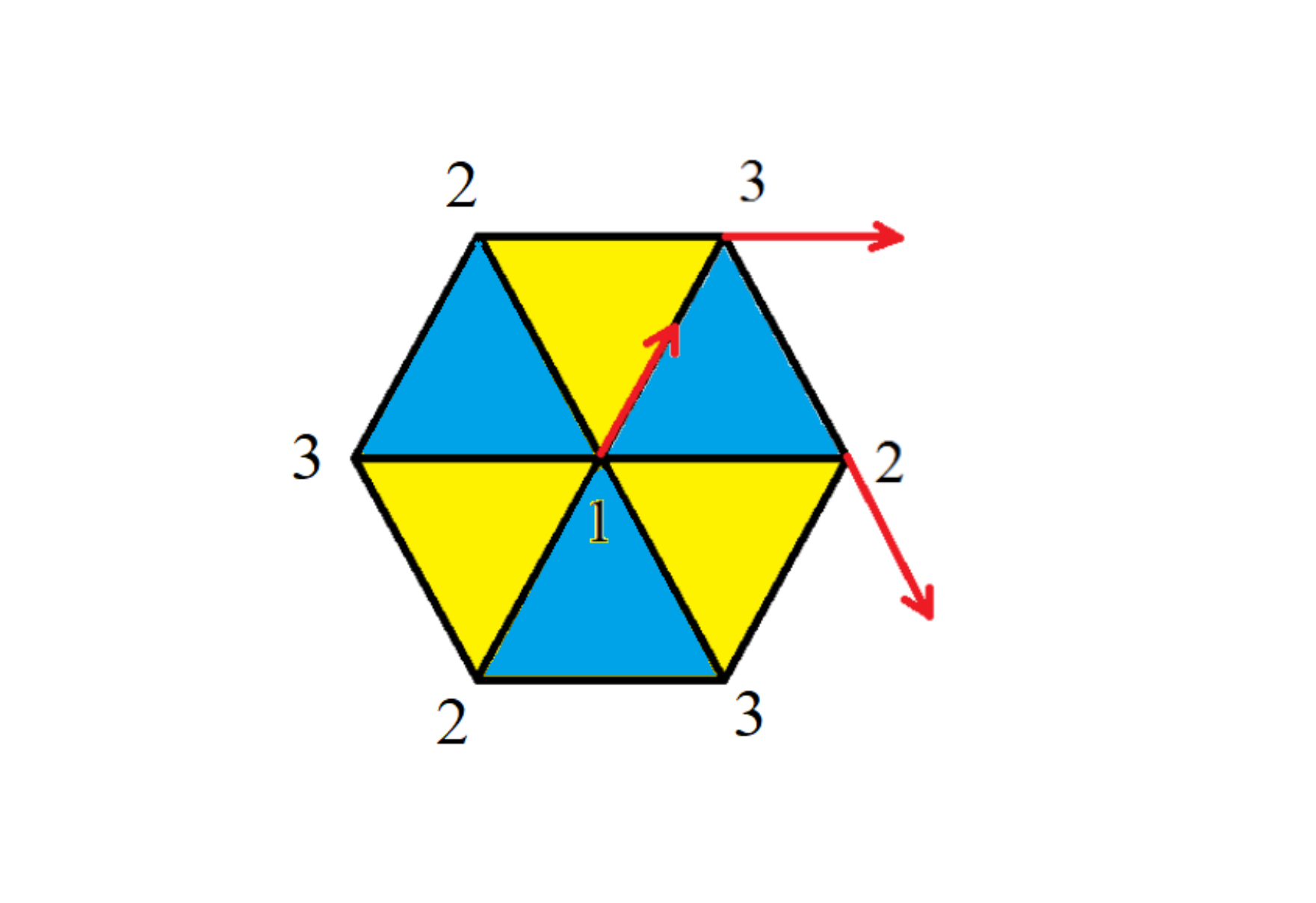}
\caption{Perpendicular $\mathbf D_{i,j}$: Ground-state spin configuration with only Dzyaloshinskii-Moriya interaction on the triangular lattice ($J=0)$ is analytically determined. One angle is 120 degrees, the other two are 60 degrees. Note that the choice of the 120-degree angle  in this figure is along the horizontal spin pair. This configuration is one GS, the other two GSs have the 120-degree angles on respectively the two diagonal spin pairs. Note also that the spin configuration is
invariant under the global spin rotation in the $xy$ plane. For convenience, the spins are decomposed into three sublattices numbered 1, 2 and 3. See text for explanation.}\label{DMGSfig}
\end{figure}

In the absence of $J$, unlike the bipartite square lattice where one can arrange the NN spins
to be perpendicular with each order in the $xy$ plane, the triangular lattice cannot fully satisfy the
DM interaction for each bond, namely with the perpendicular spins at the ends.  For this particular case of interest, we can analytically calculate the GS spin configuration as shown in the following. One considers a triangular plaquette with three spins numbered as 1, 2 and 3 embedded in the lattice.  For convenience, in a hexagonal (or triangular) lattice, we define the three sublattices as follows: consider the up-pointing triangles (there are 3 in a hexagon, see the blue triangles in Fig. \ref{DMGSfig}), for the first triangle one numbers in the counter-clockwise sense 1, 2, 3  then one does it for the other two up-pointing triangles of the hexagon, one sees that each lattice site belongs to a sublattice.
The DM energy of a plaquette is written as

\begin{eqnarray}
H_p&=& -2D[\mathbf u_{1,2}\cdot \mathbf S_1 \times \mathbf S_2 +
       \mathbf u_{2,3}\cdot \mathbf S_2 \times \mathbf S_3 +
       \mathbf u_{3,1}\cdot \mathbf S_3 \times \mathbf S_1]\nonumber\\
   &=& -2D[\sin \theta_{1,2}+\sin \theta_{2,3}+\sin \theta_{3,1}]\label{Hp}
\end{eqnarray}
where the factor 2 of the $D$ term takes into account the opposite neighbors outside the plaquette, and
where $\theta_{1,2}=\theta_2-\theta_1$ is the oriented angle between $\mathbf S_1$ and $\mathbf S_2$, etc.
Note that the $u$ vectors are in the same direction because we follow the counter-clockwise tour on the plaquette.


The minimization
of $H_p$ yields
\begin{eqnarray}
\frac{dH_{p}}{d\theta_1}&=&0=
-2D[ -\cos (\theta_{2}-\theta_{1})+\cos (\theta_{1}-\theta_{3})]\label{minimize1}\\
\frac{dH_{p}}{d\theta_2}&=&0=  -2D[ \cos (\theta_{2}-\theta_{1})- \cos (\theta_{3}-\theta_{2})]\label{minimize2}\\
\frac{dH_{p}}{d\theta_3}&=&0=  -2D[\cos (\theta_{3}-\theta_{2}) -\cos (\theta_{1}-\theta_{3})]
\label{minimize3}
\end{eqnarray}
The solutions for the above equations are
\begin{eqnarray}
\theta_{1,2}&=&\theta_{3,1} \ \ \mbox{so that} \ \ \theta_{3,2}=\theta_{3,1}+\theta_{1,2}=2\theta_{1,2}\label{gsangle1}\\
\theta_{2,3}&=&\theta_{1,2} \ \ \mbox{so that} \ \ \theta_{1,3}=\theta_{1,2}+\theta_{2,3}=2\theta_{2,3}\label{gsangle2}\\
\theta_{3,1}&=&\theta_{2,3} \ \ \mbox{so that} \ \ \theta_{2,1}=\theta_{2,3}+\theta_{3,1}=2\theta_{3,1}\label{gsangle3}
\end{eqnarray}

The solutions for the above equations are $\theta_{1,2}=\pm\theta_{1,3}$, $\theta_{2,3}=\pm\theta_{1,2}$ and
$\theta_{1,3}=\pm\theta_{2,3}$. We have to choose the correct sign in each spin pair  so as the relative angle between this  NN spin pair is not zero. Otherwise, if the relative angle is zero,  the interaction energy of such a pair yields the zero DM energy. The correct choices of sign finally give
\begin{eqnarray}
\theta_{1,2}&=&-\theta_{1,3} \ \ \mbox{so that} \ \ \theta_{2,3}=\theta_{2,1}+\theta_{1,3}=-2\theta_{1,2}\label{gsangle1}\\
\theta_{2,3}&=&\theta_{1,2} \ \ \mbox{so that} \ \ \theta_{1,3}=\theta_{1,2}+\theta_{2,3}=2\theta_{2,3}\label{gsangle2}\\
\theta_{1,3}&=&-\theta_{2,3} \ \ \mbox{so that} \ \ \theta_{2,1}=\theta_{2,3}+\theta_{3,1}=-2\theta_{1,3}\label{gsangle3}
\end{eqnarray}
These three equations, Eqs. (\ref{gsangle1})-(\ref{gsangle3}), should be solved.  We have from Eq. (\ref{minimize1})
$\cos (\theta_{1,2})=-\cos (\theta_{1,3})$. Using Eq. (\ref{gsangle3}) one obtains
\begin{equation}
\cos (2\theta_{3,1})=-\cos (\theta_{3,1})\ \ \rightarrow 2\cos^2(\theta_{3,1})+\cos(\theta_{3,1})-1=0
\end{equation}

This second-degree equation gives $\cos(\theta_{3,1})=\frac{-1\pm \sqrt{1+8}}{4}$. Only the solution with plus sign is acceptable so that  $\theta_{3,1}=\theta_{2,3}=\pi/3$. From Eq. (\ref{gsangle3}), one has  $\theta_{2,1}=2\pi/3$.
This is one solution summarized by Eq. (\ref{sol1}) below.
Note that we have taken one of them, Eq. (\ref{gsangle3}), to obtain explicit solutions for the three angles given in Eq. (\ref{sol1}). We can do the same calculation starting with Eqs. (\ref{gsangle1})-(\ref{gsangle2}) to get explicit solutions given in Eqs. (\ref{sol2})-(\ref{sol3}). We note that when we make a circular permutation of the indices of Eq. (\ref{sol1}) we get Eq. (\ref{sol2}), and a circular permutation of Eq. (\ref{sol2}) gives Eq. (\ref{sol3}).
One summarizes the three degenerate solutions below
\begin{eqnarray}
\theta_{3,1}=\theta_{2,3}=\pi/3, \ \ \theta_{2,1}=2\pi/3\label{sol1}\\
\theta_{1,2}=\theta_{3,1}=\pi/3, \ \ \theta_{3,2}=2\pi/3\label{sol2}\\
\theta_{2,3}=\theta_{1,2}=\pi/3, \ \ \theta_{1,3}=2\pi/3\label{sol3}
\end{eqnarray}

We show in Fig. \ref{DMGSfig} the spin orientations of the solution (\ref{sol1}). The GS energy is obtained by replacing the angles into Eq. (\ref{Hp}). For the three solutions, one gets the energy of the plaquette
\begin{equation}\label{GSE}
H_p= -3D\sqrt{3}
\end{equation}
We have three degenerate GSs.

\begin{figure}[h!]
\centering
\vspace{1cm}
\includegraphics[width=12cm]{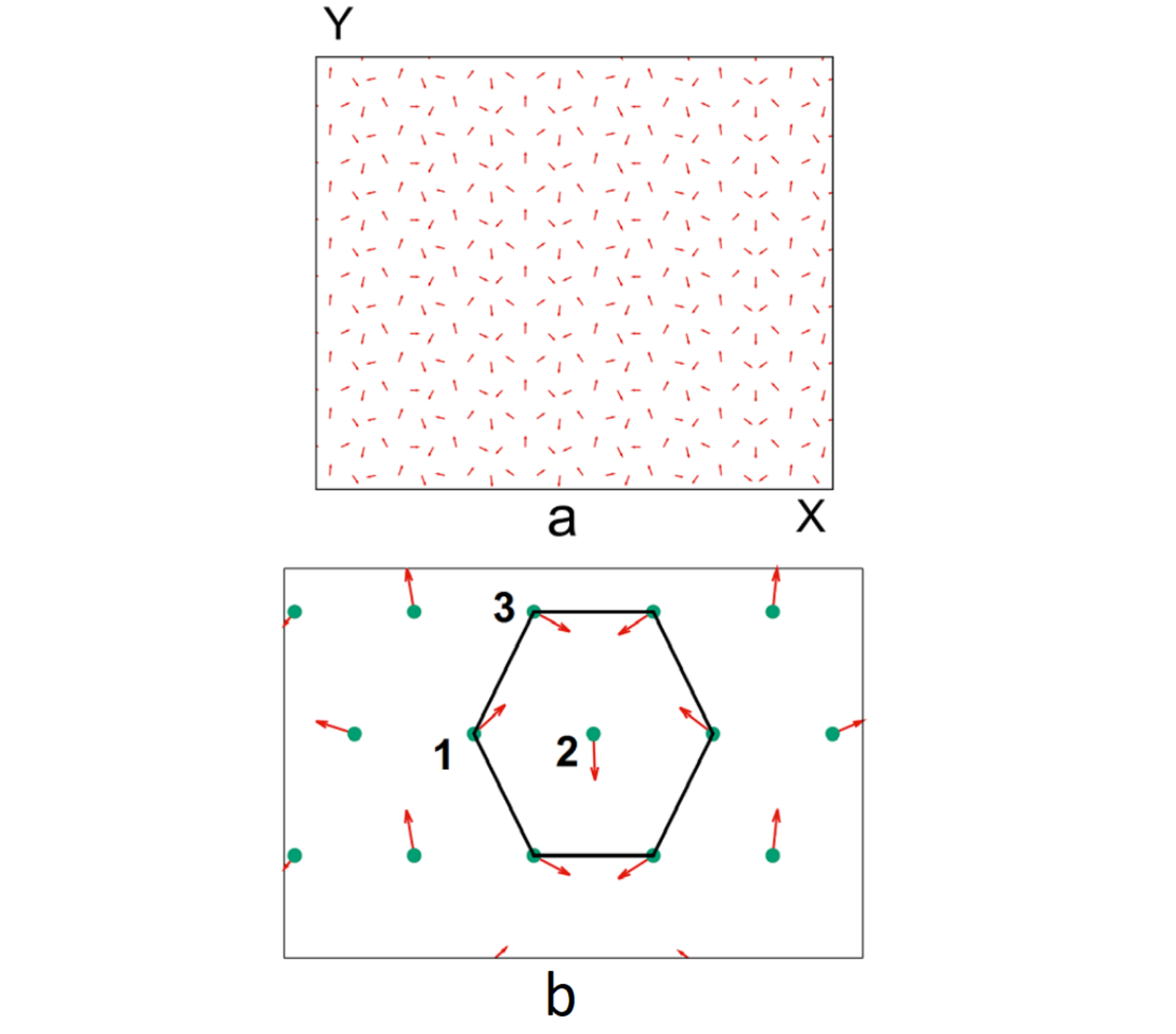}
\caption{Perpendicular $\mathbf D_{i,j}$: (a) Ground-state spin configuration with only Dzyaloshinskii-Moriya interaction on the triangular lattice ($J=0)$ obtained numerically by the steepest descent method, (b)  a zoom on a hexagonal cell, this is exactly what obtained analytically shown in Fig. \ref{DMGSfig} with a global spin rotation in the $xy$ plane: the angle of the horizontal pair (1,2) is 120 degrees, those of (2,3) and (3,1) are equal to 60 degrees.}\label{DMGSLfig}
\end{figure}

Note that this solution can be numerically obtained by
the steepest descent method described above. The result is shown in Fig. \ref{DMGSLfig} for the full lattice.
We see in the zoom that the spin configuration on a plaquette is what obtained analytically, with a global spin rotation as explained in the caption of Fig.  \ref{DMGSfig}.

As said above, to use the steepest descent method, we consider a triangular lattice of lateral dimension $L$. The total number of sites $N$ is given by $N=L \times L$. To avoid the finite size effect, we have to find the size limit beyond which the GS does not depend on the lattice size.  This is found for $L\geq 100$. Most of calculations have been performed for $L=100$.

\subsection{Ground State with both perpendicular $\mathbf D$ and $J$ in Zero Field- Spin Waves}\label{SWT}

\begin{figure}[h!]
\center
\includegraphics[width= 12cm]{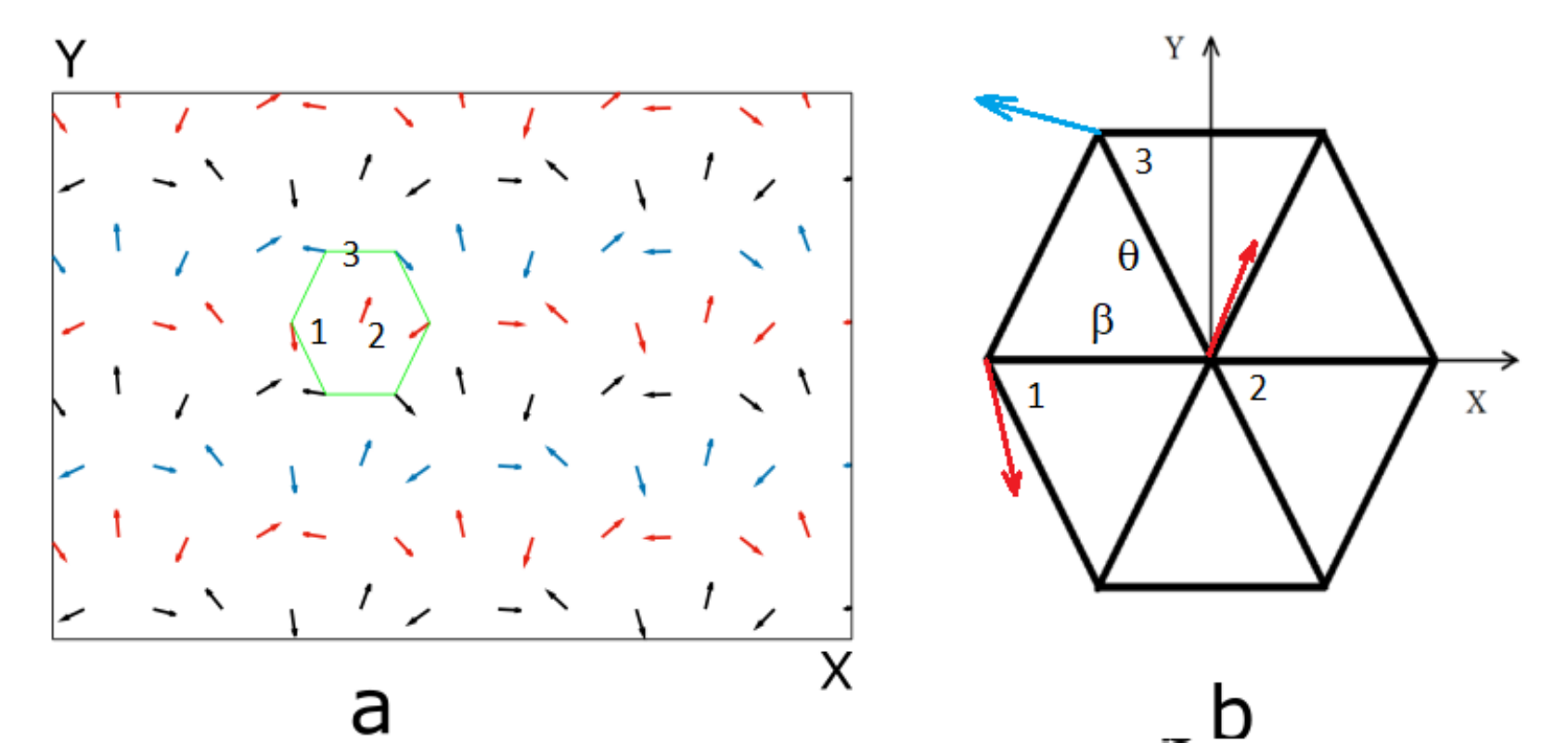}
\caption{Perpendicular $\mathbf D_{i,j}$ with antiferromagnetic $J$: (a) Ground-state spin configuration in zero field for $D = 0.5$, $J = -1$ where the angles in a hexagon are shown in (b) with $\beta = 156$ degrees for the pair (1,2) on the horizontal axis and  $\theta =  102$ degrees for the pairs (2,3) and (3,1) on the diagonals. Note that there are two other degenerate states where $\beta$ is chosen for the pair (2,3) or (3,1). }.\label{GSDJH0}
\end{figure}

When both $J$ and perpendicular $\mathbf D$ are present, a compromise is established between these competing interactions.  In zero field, the GS shows non-collinear but periodic in-plane spin configurations.  The planar spin configuration is easily understood: when $\mathbf D$ is perpendicular and without $J$, the spins are in the plane. When $J$ is antiferromagnetic without $\mathbf D$, the spins are also in the plane and form a 120-degree structure. When $\mathbf D$ and $J$ exist together the angles between NN's change but they still in the plane  in order to keep both $D$ and $J$ interactions as low as possible. An example is shown in Fig. \ref{GSDJH0} where one sees that the GS is planar and characterized by two angles  $\theta=102 $ degrees and one angle  $\beta=156$ degrees formed by three spins on a triangle plaquette.  Note that there are three degenerate states where $\beta$ is chosen for the pair (1,2) (Fig.   \ref{GSDJH0}a) or the pair (2,3) or the pair (3,1). Changing the value of $D$ will change the angle values. Changing the sign of $D$ results in a change of the sense of the chirality, but not the angle values.

In the case of perpendicular $\mathbf D_{i,j}$ in zero-field, as shown above we find the GS on a hexagon of the lattice  is defined by four identical angles
$\beta$ and two angles $\theta$ as shown in Fig. \ref{GSDJH0}. The values of $\beta$ and $\theta$ depend on the value of $D$. We take $J=-1$ (antiferromagnetic) hereafter. For $D=0.5$ we have $\beta =  156$ degrees and $\theta = 102$ degrees. For $D=0.4$ we obtain $\beta =  108$ degrees and $\theta = 144$ degrees, using $N=60\times 60$.

The periodicity of the GS allows us to calculate the SW spectrum in the following.

The model for the calculation of the SW spectrum uses quantum Heisenberg spins of magnitude $1/2$, it is given by
\begin{eqnarray}
\mathcal H=-J\sum_{\left<i,j\right>}\mathbf{S_i}\cdot\mathbf{S_j}-D \sum_{\left<i,j\right>} \mathbf u_{i,j} \cdot \mathbf{S_i} \times \mathbf{S_j} - I \sum_{\left<i,j\right>} S_i^z~S_j^z \cos\theta_{ij}
\end{eqnarray}
where $\theta_{ij}$ is the angle between $\mathbf{S_i}$ and $\mathbf{S_j}$ and the last term is an extremely small anisotropy added to stabilize the SW when the wavelength $k$ tends to zero \cite{DiepTM,Mermin}. Note that  $\mathbf u_{i,j}$ points up and down along the $z$ axis for respective two opposite neighbors.

As before, in order to calculate the SW spectrum for systems of non-collinear spin configurations, we have to use the system of local coordinates.
The  Hamiltonian becomes

\begin{equation*}
\begin{aligned}
\mathcal {H}=&-J\sum_{\left<i,j\right>} \frac{1}{4}(S_i^+S_j^++S_i^-S_j^-) (\cos\theta_{ij}-1)
+\frac{1}{4}(S_i^+S_j^-+S_i^-S_j^+) (\cos\theta_{ij}+1)\\
&+\frac{1}{2}S_j^z\sin\theta_{ij}(S_i^++S_i^-)-\frac{1}{2}\sin\theta_{ij}S_i^z(S_J^++S_j^-)+S_i^zS_j^z\cos\theta_{ij} \\
&-D\sum_{\left<i,j\right>} S_i^z S_j^z \sin\theta_{i,j} + \frac{1}{4}  \sin\theta_{i,j}( S_i ^+ S_j^+ + S_i ^+ S_j^- +
S_i ^- S_j^+)
+ \frac{1}{2}\cos\theta_{i,j}(S_i^z(S_j^+ + S_j^-) - S_j^z(S_i^+ + S_i^-))\\
&-I \sum_{\left<i,j\right>} S_i^z~S_j^z \cos\theta_{i,j}
\end{aligned}
\end{equation*}

We define the two GFs by Eqs. (\ref{green59a})-(\ref{green60}) and use
the equations of motion of these functions (\ref{eq:HGEoMG})-(\ref{eq:HGEoMF}), we obtain

\begin{equation*}
\begin{aligned}
i\hbar \frac{dG_{i,j}(t-t')}{dt}&=2<S_i^z>\delta_{i,j}\delta(t-t')- J\sum_{\left<l\right>}  <S_i^z>F_{l,j}(t-t')(\cos\theta_{i,l}-1)\\
&+<S_i^z>G_{l,j}(t-t')(\cos\theta_{i,l}+1)-2\cos\theta_{i,l}<S_l^z>G_{i,j}(t-t')\\
&+D \sum_{\left<l\right>} 2\sin\theta_{i,l} <S_i^z>F_{l,j}(t-t') - \sin\theta_{i,l} <S_i^z>( G_{l,j}(t-t') + F_{l,j}(t-t')) \\
&- 2I \sum_{\left<l\right>} \cos\theta_{i,l} <S_i^z>F_{l,j}(t-t')
\end{aligned}
\end{equation*}

\begin{equation*}
\begin{aligned}
i\hbar \frac{dF_{i,j}(t-t')}{dt}&= J\sum_{\left<l\right>}  <S_i^z>G_{l,j}(t-t')(\cos\theta_{i,l}-1)\\
&+<S_i^z>F_{l,j}(t-t')(\cos\theta_{i,l}+1)-2\cos\theta_{i,l}<S_l^z>F_{i,j}(t-t')\\
&-D \sum_{\left<l\right>} 2\sin\theta_{i,l} <S_i^z>G_{l,j}(t-t') - \sin\theta_{i,l} <S_i^z>( G_{l,j}(t-t') + F_{l,j}(t-t')) \\
&+ 2I \sum_{\left<l\right>} \cos\theta_{i,l} <S_i^z>G_{l,j}(t-t')
\end{aligned}
\end{equation*}
Note that $<S_i^z>$ is the average of the spin $i$ on its local quantization axis in the local-coordinates system (see Ref. \cite{Diep2017}).
We use now the time Fourier transforms
of the $G$ and $F$, we get

\begin{equation}
\begin{aligned}
\hbar\omega g_{i,j}&=2\mu_{i} \delta_{i,j}- J\sum_{\left<l\right>}  \mu_{i} f_{l j}
e^{-i \mathbf k\cdot (\mathbf R_{i}- \mathbf R_{l})}     (\cos\theta_{i,l}-1)\\
&+\mu_{i} g_{l j} e^{-i \mathbf k \cdot (\mathbf R_{i}-\mathbf R_{l})} (\cos\theta_{i,l}+1)-2\mu_{l}\cos\theta_{i,l} g_{i,j}\\
&-D\sum_{\left<l\right>} 2\sin\theta_{i,l} \mu_{l} g_{i,j} - \sin\theta_{i,l} \mu_{i}( g_{l,j}e^{-i \mathbf k\cdot (\mathbf R_{i}-\mathbf R_{l})} +f_{l,j}e^{-i \mathbf k\cdot (\mathbf R_{i}-\mathbf R_{l})})\\
& + 2I \sum_{\left<l\right>} \mu_{l} \cos\theta_{i,l}  g_{i,j}\label{FourierG}
\end{aligned}
\end{equation}
and
\begin{equation}
\begin{aligned}
\hbar\omega f_{i,j}&= J\sum_{\left<l\right>}  \mu_{i} g_{l j}
e^{-i\mathbf k\cdot (\mathbf R_{i}-\mathbf R_{l})}(\cos\theta_{i,l}-1)\\
&+\mu_{i} f_{l j} e^{-i \mathbf k\cdot (\mathbf R_{i}-\mathbf R_{l})} (\cos\theta_{i,l}+1)-2\mu_{l}\cos\theta_{i,l} f_{i,j}\\
&+D\sum_{\left<l\right>} 2\sin\theta_{i,l} \mu_{l} f_{i,j} - \sin\theta_{i,l} \mu_{i}( g_{l,j}e^{-i \mathbf k\cdot(\mathbf R_{i}-\mathbf R_{l})} +f_{l,j}e^{-i \mathbf k\cdot (\mathbf R_{i}-\mathbf R_{l})})\\
& - 2I \sum_{\left<l\right>} \mu_{l} \cos\theta_{i,l}  f_{i,j}\label{FourierF}
\end{aligned}
\end{equation}
where $\mu_i\equiv <S^z_i>$,  $\mathbf k$ is the wave vector in the reciprocal lattice of the triangular lattice, and $\omega$ the SW frequency.   Note that the index $z$ in $S^z_i$ is not referred to the real space direction $z$, but to the quantization axis of the spin $\mathbf S_i$.
At this stage, we have to replace $\theta_{i,j}$ by either $\beta$ or $\theta$ according on the GS spin configuration given above (see Fig. \ref{GSDJH0}).

As in the previous sections, writing the above equations under a matrix form, we have
\begin{equation}
\mathbf M \left( \hbar\omega \right) \mathbf h = \mathbf C,
\end{equation}
where $\mathbf M\left(\hbar\omega\right)$ is a square matrix of dimension
$2\times 2$, $\mathbf h$ and $\mathbf C$ are
given by
\begin{equation}
\mathbf h = \left(%
\begin{array}{c}
  g_{i,j} \\
  f_{i,j} \\
\end{array}%
\right) , \hspace{1cm}\mathbf C =\left(%
\begin{array}{c}
  2 \left< S^z_i\right>\delta_{i,j}\\
  0 \\
\end{array}%
\right) , \label{eq:HGMatrixgu}
\end{equation}
and the matrix  $\mathbf M\left( \hbar\omega\right)$ is given by
\begin{equation*}
\hspace{-2cm}
\mathbf M \left( \hbar\omega\right)= \left(%
\begin{array}{*{2}c}
 \hbar\omega +A &B \\
   -B    & \hbar\omega-A \\
\end{array}%
\right)
\end{equation*}

The nontrivial solution of $g$ and $f$ imposes the following secular equation:

\begin{equation}\label{secular}
\hspace{-2cm}
0 = \left(%
\begin{array}{*{2}c}
  \hbar\omega+A &B \\
   -B    & \hbar\omega-A \\
\end{array}%
\right)
\end{equation}
where

\begin{equation}
\begin{aligned}
A&= -J(8 \mu_{i} \cos\beta (1+I) + 4 \mu_{i} \cos\theta (1+I) - 4 \mu_{i} \gamma (\cos\beta +1) - 2 \mu_{i} \alpha (\cos\theta + 1))\\
&  -D(4 \mu_{i} \sin\beta \gamma + 2\mu_{i} \sin\theta \alpha) + D(8\mu_{i} \sin\beta  + 4\mu_{i} \sin\theta)
\end{aligned}
\end{equation}
\begin{equation}
\begin{aligned}
B&= J( 4 \mu_{i} \gamma (\cos\beta -1) + 2 \mu_{i} \alpha (\cos\theta - 1))-D(4 \gamma \mu_{i} \sin\beta + 2\mu_{i} \alpha \sin\theta)
\end{aligned}
\end{equation}
where the sum on the two NN on the $x$ axis (see Fig. \ref{GSDJH0}b) is
\begin{equation}
\sum_{l} e^{-i\mathbf k\cdot (\mathbf R_{i}-\mathbf R_{l})}= 2\cos(k_x)\equiv 2  \alpha
\end{equation}
and the sum on the four NN on the oblique directions of the hexagon (see Fig. \ref{GSDJH0}b) is
\begin{equation}
\sum_{l} e^{-i \mathbf k \cdot (\mathbf R_{i}-\mathbf R_{l})}  = 4 \cos(k_x/2) \ \cos(\sqrt{3} k_y/2)\equiv 4\gamma
\end{equation}
Solving Eq. (\ref{secular}) for each given ($k_x,k_y$) one obtains the SW frequency $\omega (k_x,ky)$:
\begin{equation}\label{SWspectrum}
(\hbar \omega)^2=A^2-B^2\ \ \rightarrow \hbar \omega=\pm \sqrt{A^2-B^2}
\end{equation}
Plotting $\omega (k_x,ky)$ in the space $(k_x,ky)$ one obtains the full SW spectrum.

The spin length $\langle S^{z}_i\rangle$ (for all $i$, by symmetry) is given by (see technical details in Ref. \cite{DiepTM}):
\begin{equation}\label{lm2}
\langle S^{z}\rangle\equiv \langle S^{z}_i\rangle=\frac{1}{2}-
   \frac{1}{\Delta}
   \int
   \int dk_xdk_z
   \sum_{i=1}^{2}\frac{Q(E_i)}
   {\mbox{e}^{E_i/k_BT}-1}
\end{equation}
where $E_i (i=1,2)=\pm \sqrt{A^2-B^2}$ are the two solutions given above, and $Q(E_i)$ is the determinant (cofactor) obtained by replacing the first column of
$\mathbf M$ by $\mathbf C$ at $E_i$.

The spin length $\langle S^{z}\rangle$ at a given $T$ is calculated self-consistently by following the method given in Refs. \cite{DiepTM,Diep2017}.

\begin{figure}[h!]
\center
\vspace{-2cm}
\includegraphics[width= 12cm]{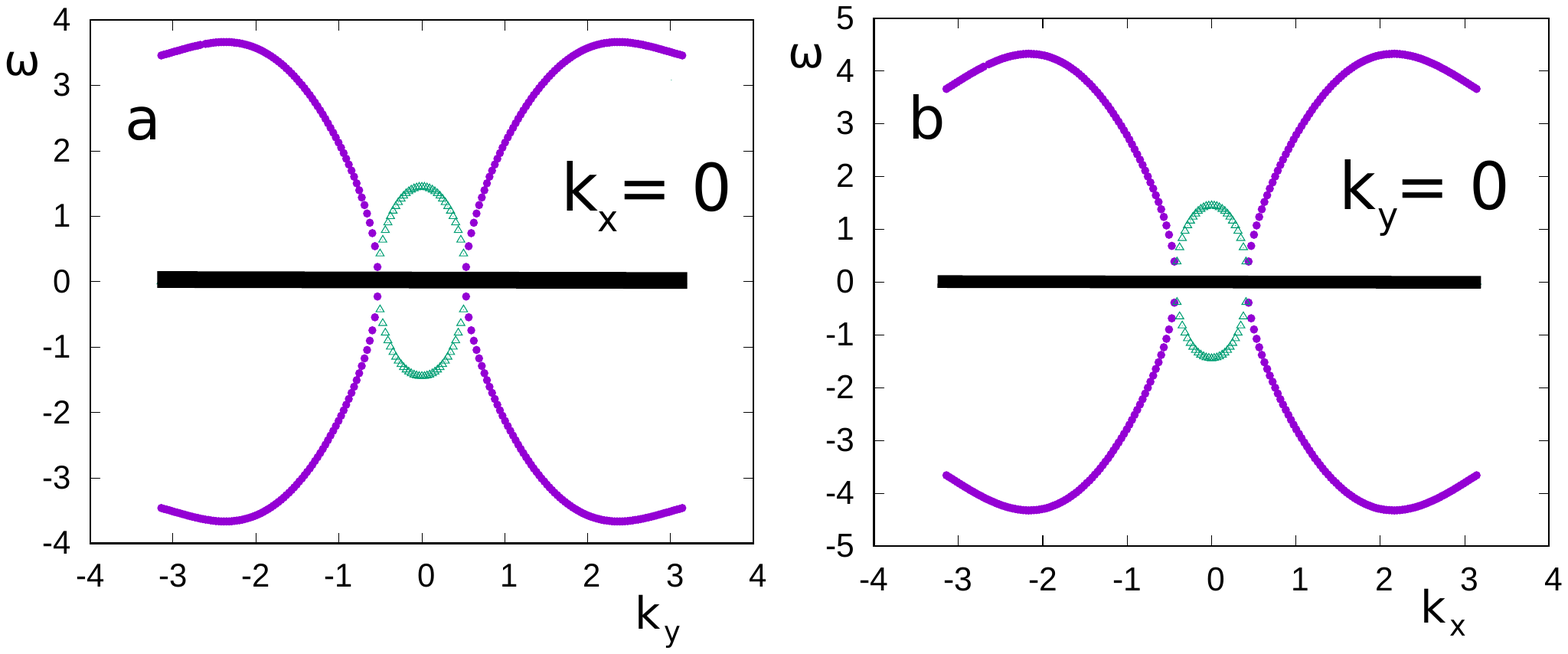}
\vspace{-7cm}
\caption{ (a) Spin-wave spectrum versus $k_y$ with $k_x=0$ at $T=0$ for $I= 0$, (b)
Spin-wave spectrum versus $k_x$ with $k_y=0$ at $T=0$ for $I= 0$. The magenta curves show the real frequency, while the green ones show the imaginary frequency. See text for comments. Parameters: $D = 0.5$, $J = -1$, $H=0$ where $\theta = 102$ degrees and  $\beta =  156$ degrees (see the spin configuration shown in Fig. \ref{GSDJH0}), $\hbar=1$.}\label{SWD1}
\end{figure}

Let us show the SW spectrum $\omega$ (taking $\hbar=1$)  for the case of $J=-1$ and $D=0.5$ in Fig. \ref{SWD1} versus $k_y$ with $k_x=0$ (Fig. \ref{SWD1}a) and versus $k_x$ for $k_y=0$ (Fig. \ref{SWD1}b).  In order to see the effect of the DM interaction alone we take the anisotropy $I=0$.
One observes here that for a range of small wave-vectors the SW frequency is imaginary.  The SW corresponding to these modes do not propagate in the system.  Why do we have this case here? The answer is that when the NN make a large angle (perpendicular NN, for example), one cannot define a wave vector in that direction. Physically, when $k$ is small the $B$ coefficient is larger than $A$ in Eq. (\ref{SWspectrum}) giving rise to imaginary $\omega$. Note that the anisotropy $I$ is contained in $A$ so that increasing $I$ for small $k$ will result in $A>B$ making $\omega$ real.

\begin{figure}[h!]
\center
\vspace{-3cm}
\includegraphics[width= 12cm]{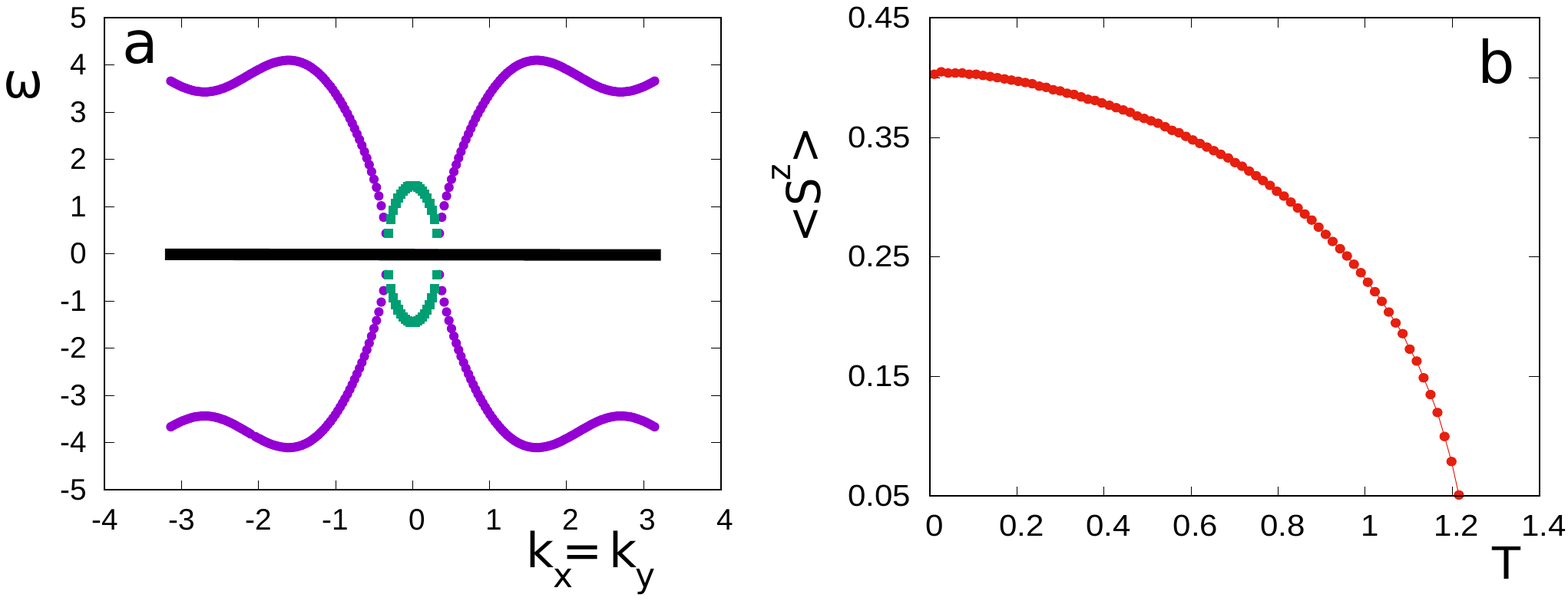}
\vspace{-7cm}
\caption{ (a) Spin-wave spectrum versus $k_x=k_y$ at $T=0$ for $I= 0$. The magenta curves show the real frequency, while the green ones show the imaginary frequency. See text for comments, (b) The spin length $S^z$ versus temperature $T$ ($k_B=1$). Parameters: $D = 0.5$, $J = -1$, $H=0$ where $\theta = 102$ degrees and  $\beta =  156$ degrees (see the spin configuration shown in Fig. \ref{GSDJH0}).}\label{SWD2}
\end{figure}

We show now in Fig. \ref{SWD2}a the spectrum along the axis $k_x=k_y$ at $T=0$ for $I= 0$. Again here the frequency is imaginary for small $k$, as in the previous figure.  The spin length $<S^z>$ along the local quantization axis is shown in Fig. \ref{SWD2}b. Several remarks are in order: i) At $T=0$, the spin length is not equal to $1/2$ as in ferromagnets because of the zero-point spin contraction due to antiferromagnetic interactions (see Ref. \cite{DiepTM}), its length is $\simeq 0.40$, quite small; ii) the magnetic ordering is destroyed at $T\simeq 1.2$.

To close the present section, we note that in the case of perpendicular $\mathbf D$ considered above, we did not observe skyrmion textures when applying a perpendicular magnetic field: all spin configurations are no more planar, making the calculation of the SW spectrum more difficult. This problem is left for a future investigation.

\section{Other systems of non-collinear ground-state spin configurations: frustrated surface in stacked triangular thin films}\label{surface}
 In this section, we study by the
GF technique the effect of a frustrated surface  on the magnetic
properties of a film composed triangular layers stacked in the $z$ direction. Each lattice site is occupied by a quantum Heisenberg spin of magnitude 1/2.  Let the in-plane surface interaction be $J_s$ which
 can be antiferromagnetic  or ferromagnetic.  The other interactions in the film are
ferromagnetic. We show in the following that the GS spin configuration is
non collinear when $J_s$ is lower than a critical value $J_s^c$. The
film surfaces are then frustrated. In the frustrated case, there
are two phase transitions, one correponds to the disordering of the two surfaces and the other to the disordering of the
interior layers. The GF results agree qualitatively with Monte Carlo simulation using the classical spins (see the original paper in Ref. \cite{NgoSurface} ).

In this section we review some ot the results given in the original paper Ref. \cite{NgoSurface}, emphasizing the SW calculation and the important results. The Hamiltonian is written as
\begin{equation}
\mathcal H=-\sum_{\left<i,j\right>}J_{i,j}\mathbf S_i\cdot\mathbf
S_j -\sum_{<i,j>} I_{i,j}S_i^z S_j^z  \label{hamil1}
\end{equation}
where the first sum is performed over the NN spin pairs $\mathbf S_i$ and  $\mathbf S_j$, the second sum over their $z$ components.  $J_{i,j}$ and $I_{i,j}$ are respectively their exchange interaction and their anisotropic one. The latter is small, taken to ensure the ordering at finite $T$ when the film thickness goes down to a few layers, without this we know that a monolayer with vector spin
models does not have a long-range ordering
at finite $T$.\cite{Mermin}

Let $J_s$ be the exchange between two NN surface spins.  We suppose that all other interactions are ferromagnetic and  equal to $J$.  We shall use $J=1$ as unit of energy  in the following.

\subsection{Ground state}

In the case where $J_s$ is ferromagnetic, the GS of the film is ferromagnetic.  When $J_s$ is antiferromagnetic, the situation becomes complicated.
We recall that for a single triangular lattice with antiferromagnetic interaction, the spins are frustrated and arranged in a 120-degree configuratrion.  \cite{DiepFSS}  This structure is modified when we turn on the ferromagnetic interaction $J$ with the beneath layer.   The competition between the non
collinear surface ordering and the ferromagnetic ordering of the bulk leads to an intermediate structure which is determined in the following.  .

\begin{figure}[htb!]
\centering
\includegraphics[width= 10cm]{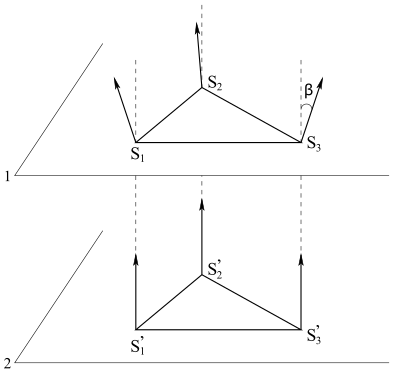}
\caption{Ground state of the film when $J_s$ is smaller than the critical value $J_s^c$. See text for description.} \label{fig:gsstruct}
\end{figure}

The GS configuration can be determined  by using
the steepest descent method described below Eq. (\ref{Ddef}).  Let us describe qualitatively the GS configuration: when $J_{s}$ is negative and  $J_s<J_{s}^c$ where $J_{s}^c(<0)$ is a critical value, the GS is formed by pulling out the
planar $120^\circ$ spin structure along the $z$ axis by an angle
$\beta$.  This is shown in  Fig.
\ref{fig:gsstruct}).

\begin{figure}[htb!]
\centering
\includegraphics[width= 10cm]{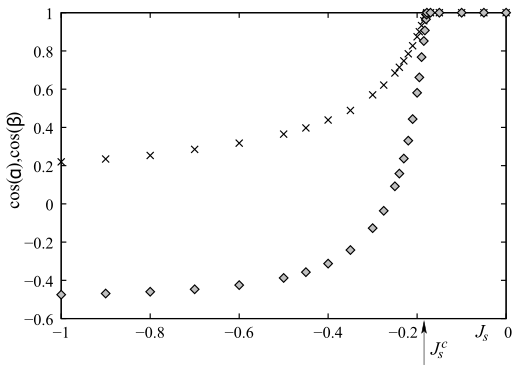}
\caption{Ground state determined by $\cos
(\alpha)$ (diamonds) and $\cos (\beta)$ (crosses) as functions of
$J_s$. Critical value of $J_s^c$ is shown by the arrow.}
\label{fig:gscos}
\end{figure}

Figure \ref{fig:gscos} shows $\cos \alpha$ and $\cos \beta$
versus $J_s$ obtained by the steepest descent method. As seen for $J_s>J_s^c$, the angles are zero, namely the GS is ferromagnetic.  The critical value $J_s^c$ is numerically found between
-0.18 and -0.19.

We show in the following that this value can be analytically calculated by
assuming the structure shown in Fig. \ref{fig:gsstruct}).
We number the spins as in that figure: $S_1$, $S_2$
and $S_3$ are the spins in the surface layer,
$S'_1$, $S'_2$ and $S'_3$ are the spins in the second layer.  The energy of the cell is
\begin{eqnarray}
H_p &=& -6\left[ J_s\left( \mathbf S_1\cdot \mathbf S_2 +\mathbf
S_2\cdot\mathbf S_3 + \mathbf S_3\cdot\mathbf S_1
\right)\right.\nonumber\\
&&+I_s\left( S^z_1S^z_2 + S^z_2S^z_3 + S^z_3S^z_1\right)\nonumber\\
&+&J\left(\mathbf S'_1\cdot \mathbf S'_2 +\mathbf
S'_2\cdot\mathbf S'_3 +\mathbf S'_3\cdot\mathbf S'_1\right)\nonumber\\
&&+I\left.\left( S'^z_1S'^z_2 + S'^z_2S'^z_3 + S'^z_3S'^z_1\right)\right] \nonumber \\
&-&2J\left( \mathbf S_1\cdot \mathbf S'_1 +\mathbf S_2\cdot\mathbf
S'_2 +\mathbf S_3\cdot\mathbf S'_3\right)\nonumber\\
&&-2I\left( S^z_1S'^z_1 + S'^z_2S'^z_2 + S^z_3S'^z_3\right),
\label{Hamilplaq}
\end{eqnarray}
We project the spins on the $xy$ plane and on the $z$ axis. One writes $\mathbf S_i=(\mathbf
S_i^{\parallel}, S_i^z)$. One observes that only surface spins have non-zero $xy$ vector
components.  Let the angle between these $xy$ components of  NN
surface spins be $\gamma_{i,j}$ which is  in fact the projection of the angle $\alpha $  on the $xy$
plane.  By symmetry, we have
\begin{equation}
\gamma_{1,2}=0,\ \gamma_{2,3}=\frac{2\pi}{3},\
\gamma_{3,1}=\frac{4\pi}{3}. \label{HSAngAlpha}
\end{equation}

The angles $\beta_i$ and $\beta'_i$ of $\mathbf S_i$ and
$\mathbf S'_i$ formed with the $z$ axis are by symmetry
$$
\left\{%
\begin{array}{c}
\beta_1=\beta_2=\beta_3=\beta,\\
\beta'_1=\beta'_2=\beta'_3=0,\\
\end{array}
\right.
$$

 The total energy
of the cell (\ref{Hamilplaq}), with $S_i = S'_i =
\frac{1}{2}$, is thus
\begin{eqnarray}
H_p&=&-\frac{9(J+I)}{2} -\frac{3(J+I)}{2}\cos\beta
-\frac{9(J_s+I_s)}{2}\cos^2\beta
\nonumber\\
&+&\frac{9J_s}{4}\sin^2\beta. \label{totEplaq}
\end{eqnarray}
The minimum of the cell energy
verifies this condition:
\begin{equation}
\frac{\partial H_p}{\partial\beta} =\left( \frac{27}{2}J_s+
9I_s\right)\cos\beta\sin\beta +\frac{3}{2}(J+I)\sin\beta \ = \ 0
\label{DerivE}
\end{equation}
One deduces
\begin{equation}
\cos\beta = -\frac{J+I}{9J_s+6I_s}. \label{GSsolu}
\end{equation}

This solution exists under the condition $-1\leq \cos\beta \leq 1$.  The critical values is determined from this condition.   For
$I=-I_s=0.1$, $J_s^c \approx -0.1889 J $ which is
in excellent agreement with the results obtained from the steepest descent method.

Now, using the GF method for such a film in the way described in the previous sections, we obtain the full Hamiltonian (\ref{hamil1}) in the
local framework:
\begin{eqnarray}
\mathcal H &=& - \sum_{<i,j>}
J_{i,j}\Bigg\{\frac{1}{4}\left(\cos\theta_{ij} -1\right)
\left(S^+_iS^+_j +S^-_iS^-_j\right)\nonumber\\
&+& \frac{1}{4}\left(\cos\theta_{ij} +1\right) \left(S^+_iS^-_j
+S^-_iS^+_j\right)\nonumber\\
&+&\frac{1}{2}\sin\theta_{ij}\left(S^+_i +S^-_i\right)S^z_j
-\frac{1}{2}\sin\theta_{ij}S^z_i\left(S^+_j
+S^-_j\right)\nonumber\\
&+&\cos\theta_{ij}S^z_iS^z_j\Bigg\}- \sum_{<i,j>}I_{i,j}S^z_iS^z_j
\label{eqHGH2}
\end{eqnarray}
where $\cos\left(\theta_{ij}\right)$ is the angle between two NN
spins.  We define the two coupled GF, and we write their equations of motions in the real space. Taking the Tyablikov's decoupling scheme to reduce
higher-order GFs, and then using the Fourier transform in the $xy$ plane we arrive at a matrix equation as in the previous section with the matrix $\mathbf M$ is defined as

\begin{equation}
\mathbf M\left(\omega\right) = \left(%
\begin{array}{ccccc}
  A^+_1    & B_1    &  D^+_1 &  D^-_1 & \cdots \\
  -B_1     & A^-_1  & -D^-_1 & -D^+_1 & \vdots \\
   \vdots  & \cdots & \cdots & \cdots &\vdots\\
  \vdots   & C^+_{N_z}   & C^-_{N_z}   & A^+_{N_z}      & B_{N_z}\\
  \cdots        & -C^-_{N_z}  & -C^+_{N_z}  & -B_{N_z}       & A^-_{N_z}\\
\end{array}%
\right), \label{HGMatrixM}
\end{equation}
where
\begin{eqnarray}
A_n^\pm &=&  \omega \pm\Big[\frac{1}{2}J_n \left< S^z_n\right>
\left(Z\gamma\right)\left(\cos\theta_{n} +1\right)\nonumber\\
&-& J_n \left< S^z_n\right>Z\cos\theta_{n} -J_{n, n+1}\left< S^z_{n+1}\right>\cos\theta_{n,n+1} \nonumber\\
&-& J_{n, n-1}\left< S^z_{n-1}\right>\cos\theta_{n,n-1} -Z
I_{n} \left< S^z_n\right>\nonumber\\
&&\ -\ I_{n,n+1}\left< S^z_{n+1}\right>-I_{n,n-1} \left<
S^z_{n-1}\right>\Big],\\
B_n &=& \frac{1}{2}J_{n}\left<
S^z_n\right>\left(\cos\theta_{n}-1\right)\left(Z\gamma\right),\\
C_n^\pm &=& \frac{1}{2}J_{n,n-1}\left<
S^z_n\right>\left(\cos\theta_{n,n-1}\pm 1\right),\\
D_n^\pm &=& \frac{1}{2}J_{n,n+1}\left<
S^z_n\right>\left(\cos\theta_{n,n+1}\pm 1\right),
\end{eqnarray}
where $Z=6$ is the in-plane coordination number, $\theta_{n,n\pm 1}$ denotes
the angle between two NN spins belonging to the adjacent layers $n$ and
$n\pm1$, while $\theta_{n}$ is the angle between two NN spins of the
layer $n$, and
$$\gamma =\left[ 2\cos \left( k_x a \right)
+ 4\cos \left( k_y a/2 \right)\cos\left( k_y
a\sqrt{3}/2\right)\right]/Z.$$

Notee that in the above coefficients,  we have used the following notations:

i) $J_n$ and $I_n$ are the in-plane interactions.  $J_n$ is equal to $J_s$ for the two surface layers and equal
to $J$ for the interior layers. All $I_n$ are taken equal to $I$.

ii) The interlayer interactions are denoted by $ J_{n,n\pm 1}$ and $ I_{n,n\pm 1}$. Note that $ J_{n,n-1}=I_{n,n-1}$=0 if
$n=1$ and $ J_{n,n+1}=I_{n,n+1}$=0 if $n=N_z$.

As described in the previous sections, the SW spectrum $\omega$ is obtained by solving det$|\mathbf M|=0$. Using $\omega$ we calculate the magnetizations layer by layer for typical values of parameters.  The results are shown in the following.


\subsection{Quantum surface phase transition}

\begin{figure}[hbt!]
\centering
\includegraphics[width= 10cm]{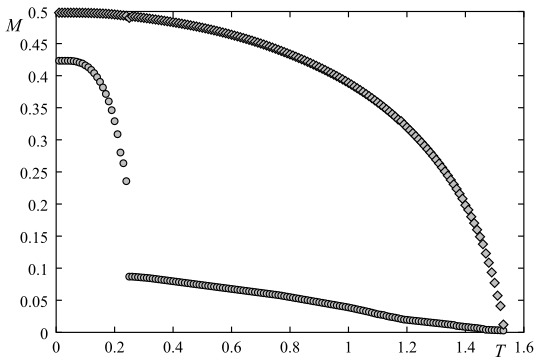}
\caption{First
two layer-magnetizations obtained by the Green function technique
vs. $T$ for $J_{s} = -0.5$ with $I=-I_s=0.1$. The surface-layer
magnetization (lower curve) is much smaller than the second-layer
one. See text for comments.} \label{fig:HGn05Ms}
\end{figure}

Let us show a typical case in the region of frustrated surface where $J_{s} = -0.5$ in Fig.
\ref{fig:HGn05Ms}. Several comments are in order:

(i) The surface magnetization is very small with respect to the magnetization of the second layer,

(ii) At $T=0$, the length of the surface spin is about 0.425 much shorter  than the spin magnitude 1/2. This is due to the antiferromagnetic interaction at the surface which causes a strong spin contraction. For the second layer, the spins are aligned ferromagnetically, their length is fully 0.5,

(iii) The surface undergoes a phase transition at $T_1\simeq 0.2557$ while the second layer remains
ordered up to $T_2\simeq 1.522$.   The system is thus
 disordered at the surface and ordered in the bulk, for temperatures between $T_1$ and $T_2$.
This partial disorder is very interesting, it gives another example of the partial disorder observed
earlier in bulk frustrated quantum spin
systems.\cite{QuartuJMMM1997,santa2}

(iv) One observes that between $T_1$ and $T_2$, the first layer has a small magnetization. This is understood by the fact that  the strong magnetization of the second layer acts as an external field on the
first layer, inducing therefore a small value of its
magnetization.

\begin{figure}[hbt!]
\centering
\includegraphics[width= 10cm]{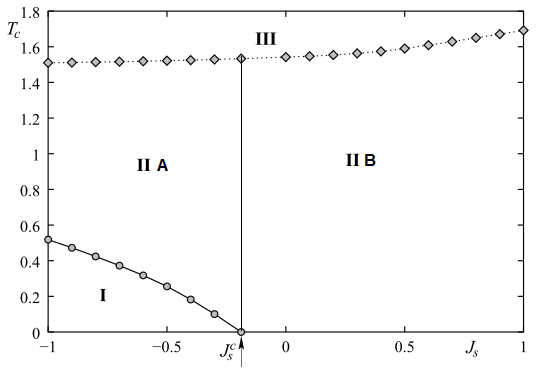}
\caption{Phase
diagram in the space ($J_{s},T$) for the quantum Heisenberg model
with $N_z=4$, $I=|I_s|=0.1$. See text for the description of
phases I to III.} \label{fig:HGDG}
\end{figure}

We plot the phase diagram in the space
$(J_s,T)$ in Fig. \ref{fig:HGDG}.  Phase I denotes the surface canted-spin state, phase IIA denotes the partially ordered phase: the surface is disordered while the bulk is ordered. Phase IIB separated from phase IIA by a vertical line issued from $J_s^c\simeq-0.19$ indicates the ferromagnettic state,  and phase III is the paramagnetic
phase.

\subsection{Classical phase transition: Monte Carlo results}
In order to compare with the quantum model shown in the previous subsection, we  consider here the classical counterpart model, namely we use the same  Hamiltonian \ref{eqn:hamil1} but with the
classical Heisenberg spin of magnitude $S=1$.  The aim  is to compare their qualitative features, in particular the question of the partial disordering at finite $T$.

We use Monte Carlo simulations for the classical model whre the film dimensions are $N\times N \times N_z$,
$N_z$ being the film thickness which is taken to be $N_z=4$ as in the quantum case shown above. We use here $N=24,
36, 48, 60$ to see the lateral finite-size effect.
Periodic boundary conditions are used in the $xy$ planes. We discard $10^6$ MC steps per spin to equilibrate the system and average physical quantities over the next $2\times 10^6$ MC steps per spin.



\begin{figure}[hbt!]
\centering
\includegraphics[width= 10cm]{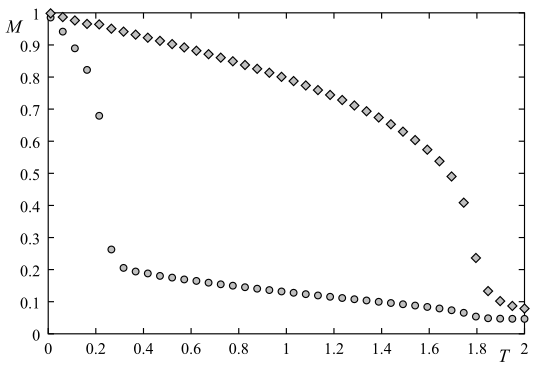}
\caption{Magnetizations of layer 1 (circles) and layer 2
(diamonds) versus temperature $T$ in unit of $J/k_B$ for
$J_s=-0.5$ with $I=-I_s=0.1$.} \label{fig:HSn05Ms}
\end{figure}

We show in Fig. \ref{fig:HSn05Ms} the result obtained in the same frustrated case as  in the quantum case shown  above, namely
$J_s=-0.5$.  we see that
the surface magnetization falls at $T_1\simeq 0.25$ while the second-layer magnetization stays ordered up to $T_2\simeq 1.8$. This surface disordering at low $T$ is similar to the quantum case. Between $T_1$ and $T_2$ the system is partially disordered.

\begin{figure}[hbt!]
\centering
\includegraphics[width= 10cm]{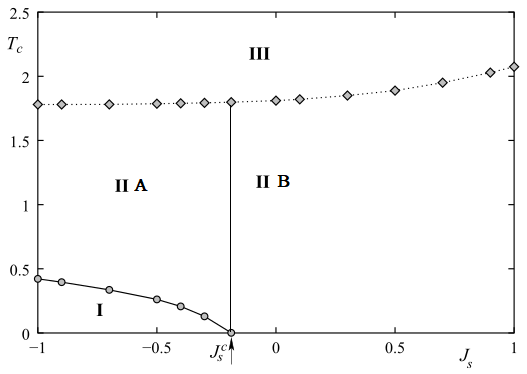}
\caption{Phase
diagram for the classical Heisenberg spin
using the same parameters as in the quantum case, i. e.  $N_z=4$, $I=|I_s|=0.1$. The definitions of phases I to III have been given in the caption of Fig. \ref{fig:HGDG} .} \label{fig:HSDG}
\end{figure}

Figure \ref{fig:HSDG} shows the phase diagram obtained in the space
$(J_s,T)$.  It is interesting to note that the classical phase diagram shown here has the same feature as the quantum phase diagram displayed
in Fig. \ref{fig:HGDG}.  The difference in the values of the transition temperatures is due to the difference of quantum and classical
spins.

To close this review, we should mention a few works works where SW in the regime of non-collinear spin configurations  have been studied: the frustration effects in antiferromagnetic face-centered
cubic Heisenberg films have been studied in Ref. \cite{NgoSurface2}, a frustrated ferrimagnet  in Ref. \cite{QuartuJMMM1997} and a quantum frustrated spin system in Ref. \cite{santa2}. These results are not reviewed here to limit the paper's length. The reader is referred to those works for details.

\section{Concluding remarks}\label{concl}

As said in the Introduction, the self-consistent Green's function theory is the only one which allows to calculate the SW dispersion relation in the case of non-collinear spin configurations, in two- and three dimensions,  as well as in thin films. The non-collinear spin configurations are due to the existence of competing interactions in the system, to the geometry frustration such as in the antiferromagnetic triangular lattice, or to the competition between ferromagnetic and/or antiferromagnetic interactions with the Dzyaloshinskii-Moriya interaction. We have shown that without  an applied magnetic field, the GS spin configuration is non collinear but periodic in space.  We have in most cases analytically calculated them. We have checked them by using the iterative numerical minimization of the local energy (the so-called steepest-descent method).  The agreement between the analytical method and the numerical energy minimization is excellent.  The determination of the GS is necessary because we need them to calculate the SW spectrum: SW are elementary excitations of the GS when $T$ increases.

The double-fold purpose of this review is to show the method and the interest of its results.  We have reviewed a selected number of works according to their interest of the community:  helimagnets,  materials with the Dzyaloshinskii-Moriya interaction, and the surface effects in thin magnetic films. The  Dzyaloshinskii-Moriya interaction gives rise not only a chiral order but also the formation of skyrmions in an applied magnetic field.  The surface effects in helimagnets and in films with a frustrated surface give rise to the reconstruction of surface spin structure and many striking features due to quantum fluctuations at low $T$  such as the zero-point spin contraction and the magnetization crossover). We have also seen above the surface becomes disordered at a low $T$ while the bulk remains ordered up to a high $T$. This coexistence of bulk order and surface disorder in a temperature region is also found in several frustrated systems \cite{DiepFSS}.

To conclude, we say that the Green's function theory for non-collinear spin  systems is laborious, but it is worth to use it to get results with clear physical mechanisms lying behind observed phenomena in frustrated spin systems.

\acknowledgments{The author thanks his former doctorate students Drs. R. Quartu, C. Santamaria, V. T. Ngo, S. El Hog, A. Bailly-Reyre and I. F. Sharafullin for the collaborative works presented in this review.}

\end{document}